\documentclass[sigconf]{acmart}
\AtBeginDocument{%
  }

\copyrightyear{2026}
\acmYear{2026}
\setcopyright{cc}
\setcctype{by}
\acmConference[CHI '26]{Proceedings of the 2026 CHI Conference on Human Factors in Computing Systems}{April 13--17, 2026}{Barcelona, Spain}
\acmBooktitle{Proceedings of the 2026 CHI Conference on Human Factors in Computing Systems (CHI '26), April 13--17, 2026, Barcelona, Spain}
\acmPrice{}
\acmDOI{10.1145/3772318.3791098}
\acmISBN{979-8-4007-2278-3/2026/04}

\acmSubmissionID{3093}

\usepackage{siunitx}
\usepackage{caption}
\usepackage{subcaption}

\begin{document}


\title{Do We Need Subsidiarity in Software?}

\author{Louisa Conwill}
\email{lconwill@nd.edu}
\orcid{0009-0001-7116-266X}
\affiliation{%
  \institution{Computer Science and Engineering}
  \institution{University of Notre Dame}
  \city{Notre Dame}
  \state{Indiana}
  \country{USA}
}

\author{Megan Levis Scheirer}
\email{mlevis@nd.edu}
\orcid{0000-0002-3328-2732}
\affiliation{%
  \institution{Institute for Social Concerns}
  \institution{University of Notre Dame}
  \city{Notre Dame}
  \state{Indiana}
  \country{USA}\\
  \institution{Computer Science and Engineering}
  \institution{University of Notre Dame}
  \city{Notre Dame}
  \state{Indiana}
  \country{USA}
}

\author{Walter J. Scheirer}
\email{wscheire@nd.edu}
\orcid{0000-0001-9649-8074}
\affiliation{%
\institution{Computer Science and Engineering}
  \institution{University of Notre Dame}
  \city{Notre Dame}
  \state{Indiana}
  \country{USA}
}

\renewcommand{\shortauthors}{Conwill et al.}

\begin{abstract}
Subsidiarity is a principle of social organization that promotes human dignity and resists over-centralization by balancing personal autonomy with intervention from higher authorities only when necessary. Thus it is a relevant, but not previously explored, critical lens for discerning the tradeoffs between complete user control of software and surrendering control to “big tech” for convenience, as is common in surveillance capitalism. Our study explores data privacy through the lens of subsidiarity: we employ a multi-method approach of data flow monitoring and user interviews to determine the level of control different everyday technologies currently operate at, and the level of control everyday computer users think is necessary. We found that chat platforms like Slack and Discord violate subsidiarity the most. Our work provides insight into when users are willing to surrender privacy for convenience and demonstrates how subsidiarity can inform designs that promote human flourishing.
\end{abstract}


\maketitle


\section{Introduction}
Large tech companies loom over our day-to-day lives. Attempting to cut them out of one's life can make everyday tasks significantly more difficult, or even impossible~\cite{hill_2019_life_without}. Yet, there is a growing concern with these companies' behaviors, including their strong concentrations of economic power and their influence on society~\cite{hill_2019_life_without}. In comparison to the computer systems of the 1990s where users had more control over their devices, the typical computer user today has surrendered their ability to fully control their devices and online experiences. One major way computer users have surrendered their control is through \textit{surveillance capitalism}: the economic system employed by many tech companies offering ``free'' services that collect user data to not only predict but also influence and modify behavior~\cite{zuboff}. In addition to corporate surveillance, some users may fear that increasing tech centralization enables government surveillance~\cite{greenberg_newman_2024_wired}. Surveillance aside, other users may feel that this surrender of control limits the customization and flexibility of their machines, finding that operating systems like Windows force or strongly nudge their users to use other Microsoft products such as the Bing search engine, the Edge browser, and the Copilot AI service.~\cite{pewdiepie_linux}.

But this corporate centralization also has its benefits. Data collection from large tech companies allows for better understanding of user preferences to improve the product and provide more relevant recommendation algorithms. Data collected for advertising allows for many online services to be offered for free to all users. Many users would prefer the company developing their laptop make the design and customization decisions for them rather than putting the burden of customization on the user. ``Single sign on'' services offered by companies like Google and Meta provide users the convenience of not having to remember a username or password for many different platforms and applications. While some computer users are increasingly concerned with centralization, there may certainly be others that would not want to surrender its conveniences.

Before modern technology existed as it does today, a principle of social organization emerged as a response to increasing centralization of economic power caused by the technological advancements of the Industrial Revolution. This principle, called the principle of \textit{subsidiarity}, articulates how to balance the importance of the autonomy of individuals with the potential necessity of action from higher levels of power to achieve the common good. Subsidiarity states that decisions or control should occur at the lowest level \textit{possible} and the highest level \textit{necessary}. Importantly, control should be at the lowest level by default, and should only move to a higher level with good justification. As we approach the tension between the concerns and conveniences of centralized technological power today, the principle of subsidiarity can help us navigate how to balance the importance of digital autonomy with the conveniences of centralized control from big tech companies.

We apply subsidiarity in a technological context in the following way: \textit{control of digital spaces should be made at the lowest level possible and the highest level necessary.} This understanding of subsidiarity encompasses all of the kinds of control a user can have over their device: from customizing one's desktop experience through methods akin to ``Linux Ricing''~\cite{linux_ricing} to content moderation on a decentralized social media platform. While there are many different ways a user can hold or surrender control in the context of today's digital technologies, we choose to focus on one particular application of subsidiarity in this paper: subsidiarity in terms of data collection. In this way, we evaluate the tradeoffs between privacy and convenience called into question with surveillance capitalism today.

Applying the lens of subsidiarity to today's technologies requires an understanding of both what level of control the technologies are operating at and what level of control users find is \textit{necessary} for the effective usage of different platforms. We articulate our questions for investigation through the following research questions:
\begin{itemize}
    \item \textbf{RQ1:} \textit{To what extent is the software we use every day collecting data beyond expected or necessary usage? In other words, what level of control is the software currently operating at?}
    \item \textbf{RQ2:} \textit{What do users think is the necessary level of data control for various types of software? What is the justification?}
    \item \textbf{RQ3:} \textit{Is there a mismatch between how the software is behaving and the level of control the users desire from it? In other words, is the software abiding by or violating subsidiarity?}
\end{itemize}

We answer \textbf{RQ1} through data flow monitoring of various operating systems, browsers, email applications, search engines, social media, and chat applications, both while running idle and through regular usage. The results of our data flow monitoring led us to determine if each of these software platforms operates at a lower, more user-centered level of control, or at a higher, more platform-centered level of control. We answer \textbf{RQ2} through semi-structured interviews, asking everyday computer users what level of control they find necessary for each of these application types and why. Finally, we answer \textbf{RQ3} by comparing the results of the data flow monitoring and interviews.

We found that while our participants primarily want their data controlled at the user level, they are more willing to have it controlled at a platform level for browsers, search engines, and social media. Among the software we studied, there are generally options that align with user desires for level of control. We found the biggest violation of subsidiarity in the chat applications we studied: our participants desired the greatest level of privacy on chat applications, yet the ones we studied engaged in data sharing. Our interviews also provided the justifications for when it could be necessary to push control to a higher level, and revealed a desire for increased control at the community level: a level which lies higher than the user level but lower than the platform level.

The contributions of this paper are as follows:
\begin{itemize}
    \item An argument for why the principle of subsidiarity is a helpful frame for considering ethical questions in computing
    \item The empirical contributions of our data flow analysis, which provide insight into how various applications (including operating systems and software applications not well-studied in previous literature) could be transmitting our personal data
    \item The empirical contributions of our user interviews, which provide insight into what level of data control computer users view as necessary across different types of software applications and why
    \item An analysis of our chosen software platforms through the lens of subsidiarity. This provides insight into the gaps and opportunities in either amplifying the benefits of centralizing or minimizing risks to empower users across different types of software platforms. It additionally models how other technologies could be evaluated in light of subsidiarity
\end{itemize}

Beyond furthering the conversation regarding the ethics of data privacy, we hope this paper will demonstrate the principle of subsidiarity as an effective critical lens through which to promote user flourishing in technology design. The principle of subsidiarity can help designers concerned with user well-being promote user autonomy as much as possible, while also being able to discern and justify instances of higher-level control for the benefit of the user.

\section{Related Work}

We situate our research within a number of areas of prior work. First, we look at previous work in human-computer interaction that engages with the principle of subsidiarity or similar frameworks. Next, we consider previous research in data collection, sharing, and tracking, especially in the contexts of devices and browsers. Finally, we consider prior research of user perceptions and attitudes towards data privacy. In all cases, we highlight the unique contributions of our work.

\subsection{The Principle of Subsidiarity and Similar Frameworks in HCI and Beyond}
Two prior works in HCI have considered the principle of subsidiarity explicitly. Both demonstrated subsidiarity as an effective principle for re-imagining more ethical social media. First, Hasinoff and Schneider show subsidiarity can help improve the balance between context and scale in content moderation~\cite{hasinoff2022scalability}. We build on this work by bringing subsidiarity into a new domain and through conducting empirical studies. In their work on common good-oriented design patterns for social media, Conwill et al. argued that designing for the principle of subsidiarity would encourage more respectful and authentic conversations, and help to resist exploitative mechanisms from tech companies~\cite{conwill_design_patterns}. They also note that subsidiarity entails allowing users to customize their experiences more, giving users greater control over the content displayed in their feeds and the amount of notifications they receive. In this work we expand the notion of control into the realm of data privacy.

The principle of subsidiarity has also been proposed as a critical lens for LLM development. Specifically, a recent issue of the \textit{Red Hat Research Quarterly} proposed that subsidiarity should be a guiding principle for open-source AI development~\cite{Craven2025_OpenSourceAI}. Amid debates of how ``open'' open-source models should be, the authors propose that subsidiarity can help determine how much access different communities have to model components based on how much control of the model they need for their use case~\cite{Craven2025_OpenSourceAI}. In this way, subsidiarity is upheld when control is given to the most localized community that is able to effectively make their own decisions.

Related to subsidiarity in HCI is the Anarchist HCI movement~\cite{keyes2019human}. Social anarchists hold that ``human dignity is greatest when human autonomy is greatest'' and that communities are the most appropriate decision-makers and loci of power~\cite{keyes2019human}. In this we see resonance with subsidiarity, which also holds that autonomy supports human dignity and that lower levels of decision-making are more just. An even stronger connection between anarchy and subsidiarity lies in how subsidiarity can inform understandings of anarchy. As an example, the Catholic Worker Movement, an anarchist Catholic social justice movement, uses the principle of subsidiarity to define its understanding of anarchy~\cite{linder2024wherein}. This connection illustrates how our use of subsidiarity as a critical lens for HCI extends previous work in anarchist HCI. Previous contributions within anarchist HCI proposed how anarchism can influence HCI as a discipline, offering ways to re-think the contributions of our works and how HCI researchers relate to their study participants and the broader HCI research community~\cite{keyes2019human}. In contrast, our work considers how subsidiarity can influence the design and use of the computer technologies themselves.

\subsection{Data Sharing and Tracking at the Device Level and Browser Level}\label{sec:sharing}
In this paper, we investigate to what extent the platforms we use every day send and receive data beyond expected or necessary usage. Here we highlight prior work that has investigated data tracking on devices and browsers, and discuss how our work builds on this prior work.

Prior work in both HCI and other computer science disciplines has investigated privacy and data tracking on different types of devices. These studies have found that various types of devices --- fertility trackers~\cite{hudig2025intimate}, Apple mobile devices~\cite{bourdoucen2024privacy}, and Android devices~\cite{gamba2020analysis} --- access and share data in ways unexpected to the user. This data sharing spans countries and organizations, including to advertising and tracking services~\cite{hudig2025intimate, gamba2020analysis}. We extend this prior work both by investigating the data flows from different operating systems, specifically Windows and Ubuntu, and analyzing our findings through the critical lens of the principle of subsidiarity. We draw methodologically from the work of Hudig and Singh~\cite{hudig2025intimate}, incorporating both data flow monitoring and user interviews in our work.

At the browser level, data sharing and tracking is extremely common: the first third-party tracker emerged around 1996, and websites have contacted an increasing amount of third-party trackers over time~\cite{lerner2016internet, krishnamurthy2009privacy}. Third party trackers have begun to exploit first-party cookies for tracking ~\cite{munir2023cookiegraph, chen2021cookie} and use a technique called ``fingerprinting'' to identify individual users based on browser information~\cite{eckersley2010unique}. Although it is known that browsers engage in data sharing and tracking, we build on this prior work by considering this in light of the principle of subsidiarity. 

Google, Facebook, Microsoft, and Twitter are the most dominant third-party trackers, especially in Western countries, with Google being the significantly biggest offender~\cite{schelter2016tracking, gill2013advertising, roesner2012detecting, lerner2016internet, karaj2018whotracks, englehardt2016online, krishnamurthy2009privacy, cassel2022omnicrawl}. More recently, TikTok has become a major third-party tracker~\cite{munir2023cookiegraph}. We will rely on these findings to help determine which platforms are ``known data harvesters,'' which is one piece of our determining the level of control of an application.

Mainstream browsers experience 70\% more third-party tracking and advertising requests on average compared to privacy-focused browsers and experience fingerprinting on significantly more websites~\cite{cassel2022omnicrawl}. As such, we include one privacy-focused browser, Brave, in our investigation.

We extend this prior research first by studying Windows and Ubuntu, which do not appear to be studied in the previous literature. Second, a number of the web applications we investigated, especially BlueSky, Discord, and Slack, do not appear in the previous literature.

\subsection{User Perceptions of Privacy}
In this paper we additionally investigate user desires for the appropriate level of control --- whether this is complete control by the user or control by the software platform --- in relation to their personal data. Prior work has demonstrated that user willingness to share their data is \textit{contextual}: it is dependent on who the data is being shared with and why, and the sensitivity of the data being shared~\cite{hudig2025intimate}. In this paper we add another layer of context, investigating user desires for level of control dependent on software type and the context in which the software is used.

Users also have differing opinions about data sharing. While a majority of users have negative feelings towards data tracking but don't take any action to minimize it~\cite{coopamootoo2022feel}, in some cases, users make different choices regarding data privacy when offered greater transparency~\cite{liccardi2014no, van2017better}. People find targeted advertising to be simultaneously useful and privacy-invasive~\cite{ur2012smart} and have divergent opinions about the tradeoffs between the harms and conveniences of targeted advertising~\cite{weinshel2019oh}. Additionally, while most people find online tracking to be creepy, how people define creepy vastly differs~\cite{reitinger2024does}. These findings help to motivate the purpose of our interviews: it is not necessarily sufficient to say that perfect subsidiarity in our technologies would mean to always put control at the level of the user, when users acknowledge the benefits of surrendering their data.

\section{Theory: The Principle of Subsidiarity}\label{sec:theory}
In this section we will further explain the principle of subsidiarity, outline its history and use, and argue why we believe it is an effective critical lens through which to envision a better relationship with technology in the future.

The principle of subsidiarity states that local governing bodies should be empowered to make decisions for themselves, and central authorities should only perform tasks that cannot be done at a local level~\cite{compendium}. The principle of subsidiarity promotes human dignity through a prioritization of human agency and autonomy: by encouraging decisions to be made at the lowest level possible, individuals and communities are able to make decisions about matters that affect them most, fostering a sense of responsibility and participation~\cite{BenedictXVI_CaritasInVeritate}. However, subsidiarity is not libertarianism: an important element of subsidiarity is acknowledging the importance of higher levels of intervention when necessary and with appropriate justification~\cite{compendium}.

Historically, the principle of subsidiarity is perhaps most prominently associated with the Christian Democratic political ideology, having its roots in both Catholic Social Teaching~\cite{compendium} and the neo-Calvinist teaching of sphere sovereignty~\cite{subsidiarity_wikipedia}. However, its resonance goes beyond Christian Democrats alone. For example, today subsidiarity may be most well-known as a general principle of European Union law~\cite{subsidiarity_wikipedia}. Importantly, the principle of subsidiarity can resonate with both the political left and the political right: this broad resonance makes it worthy of our consideration. The political left can find resonance with the principle of subsidiarity because it upholds participatory government, resists centralized economic and political power, and can support self-determination for historically marginalized groups. Subsidiarity can resonate with the political right because it can justify limited government and supports giving primacy to traditional institutions like the family in addressing moral questions. 

While typically associated with questions of political organization, subsidiarity can apply in other contexts. For example, in the workplace, a boss can empower her employees through the principle of subsidiarity by avoiding micro-managing. In this way, the gifts and talents of the employees are honored by giving the employee a greater ability to contribute freely. 

Subsidiarity can offer much in the space of ethical technology design as well. Importantly, Catholic Social Teaching --- the Catholic Church's doctrine about social justice and the primary framework through which subsidiarity emerged --- began in response to the societal concerns of the Industrial Revolution~\cite{compendium}. Specifically, subsidiarity emerged as a response to increasing centralization of economic power~\cite{compendium}, a trend we see repeating today with ``big tech.'' While systems like surveillance capitalism undermine personal autonomy, subsidiarity aims to reclaim it. Yet, because subsidiarity guides the discernment of when intervention from higher levels of power is necessary, it can help to tease out the tension between user desires for privacy and the convenience of giving up their data, and help to clarify when data collection is appropriate or necessary.

As mentioned in the introduction, subsidiarity in a technical context means \textit{control of digital spaces should be made at the lowest level possible and the highest level necessary.} In the specific application to data collection, subsidiarity means that \textit{data collection should occur at the lowest level possible and highest level necessary for effective use of the application}. We identify the levels of control we consider in this paper from lowest to highest as follows: \textit{user}, \textit{community}, and \textit{platform}. Control at the user level means that data collection should be transparent and require informed consent to be collected and used in ways not immediately obvious from the primary function of the application. Control at the platform level means that companies can collect our data how they please, typically in exchange for the conveniences afforded by them using and/or sharing our data. Finally, control at the community level pertains to communities, employers, organizations, or special interest groups having control of their participants' data, typically for the greater benefit of that community. This may include an employer providing software to its employees, a high school providing Chromebooks to its students through the Google for Education program, or a community-run online forum for a special interest or hobby, among other examples.

\section{Methodology}
In light of the above, we aim to determine if common everyday technologies abide by the principle of subsidiarity. First, we investigate the level of control that different operating systems, browsers, and applications available today operate at. Second, we interview computer users to determine the level of control they believe is necessary for these different types of software platforms. Considering these findings together, we determine if the software platforms we investigated abide by the principle of subsidiarity.

Prior work has highlighted that security and privacy concerns vary across different stages of the data life cycle: data collection, usage, and retention~\cite{ali2025understanding}. In our methodology we focus on subsidiarity in relation to data collection. We  focus on data collection because it is both the gateway to the other stages and also where we find the user has the most potential for control in the current technical landscape: a user can, to some extent, control if their data is collected through choosing to use or reject particular platforms that are known to collect user data. However, once data is collected it is less straightforward for the user (at least in the current technical landscape) to control how their data will be used or retained by the platform. It is also the most straightforward stage for us to study: we can monitor if platforms are collecting data but once data is collected we cannot know for sure how they are using it. Additionally, even if platforms provide mechanisms for users to ``delete'' their data, in some cases this data may actually be retained by the platform~\cite{ng2019alexa}, making the study of data retention more difficult.

To achieve these aims we employ a multi-method approach. First, we examine network transmissions during the usage of our selected operating systems, browsers, and applications. From this, we determine if the software in question is operating at a lower (more user-centered) or higher (more platform-centered) level of control based on data flow patterns and pre-existing knowledge about company data sharing practices. Next, we interview computer users about their desires for data control in different types of applications. Finally, we compare user desires for level of control with the actual level of control the applications are operating at. From the interviews we also collect the necessary justifications for moving up in levels of control. Through this, we determine if the software we analyzed is abiding by the principle of subsidiarity: if it is operating at the level of control deemed necessary by the user interviews, then that software abides by subsidiarity.

\subsection{Data Flow Monitoring}
We monitored the network transmissions of different types of everyday software platforms in order to infer the extent to which these platforms are collecting and sharing data. Although it is impossible to get a full picture of what is going on, our aim was to get as much of a sense as we could of the entities being contacted and the potential endpoints where user data could be being transmitted. To do so, we used the network analysis tool Wireshark~\cite{wireshark} to capture the network activity while using different software platforms. We investigated the network activity of multiple operating systems and browsers on their own, as well as various web applications, run in a particular browser on a particular operating system.

\subsubsection{Operating System and Browser Selection}
We wanted to compare operating systems and browsers that are more conventional --- i.e., owned by large tech companies, who do not make many promises about privacy --- to operating systems and browsers that are open-source and/or claim to better protect user privacy. Thus, we selected the operating systems Microsoft Windows and Ubuntu Linux, and the browsers Google Chrome, Firefox, and Brave~\cite{brave}. Brave has been demonstrated in the literature to be the most effective browser at reducing third-party tracking and advertising requests~\cite{cassel2022omnicrawl}. 

\subsubsection{Application Selection}
We focused on everyday Internet activities, especially those that tend to occur on ad-focused platforms. We considered four categories of activities: \textit{email}, \textit{search}, \textit{social networking}, and \textit{chat}. Within these categories we chose:
\begin{itemize}
    \item \textit{Email:} Gmail (personal), Gmail (university sponsored), Proton Mail
    \item \textit{Search:} Google Search, Duck Duck Go, Brave Search
    \item \textit{Social Networking:} Instagram, TikTok, YouTube, X, BlueSky
    \item \textit{Chat:} Discord, Slack
\end{itemize}

\subsubsection{Data Collection}
We performed our data flow monitoring experiments on two identical Dell laptops. We used the default Windows installation on one laptop and installed Ubuntu Linux on the other. All experiments were conducted on our university's (a large research university in the Midwestern United States) guest Wi-Fi network.

\textbf{First}, we collected data on the actions of the operating systems on their own with no applications (other than Wireshark) running. This first and foremost enabled us to determine which endpoints the operating systems were communicating with and where application data was being sent unexpectedly (i.e., without any actions by the user). Secondly, this allowed us to collect a set of known IP addresses that are communicated with by the operating system. Because the browser and web application experiments inherently have to be run on an operating system, this allowed us to better filter out traffic in those experiments that was likely due to background traffic from the operating system and not the software in question. This practice of filtering out background traffic is consistent with other similar studies.~\cite{hudig2025intimate}. To do this experiment, we ran Wireshark for one hour as the Windows and Ubuntu machines sat idle.

\textbf{Second}, we collected data on the actions of the browsers with no websites navigated to. This again served the dual purpose of understanding which endpoints the browsers are contacting and sending application data to, and also determining browser background traffic. To run these experiments, we let each browser sit open and idle for one hour as we tracked the network traffic on Wireshark. We conducted these experiments on both the Windows and Ubuntu machines.

\textbf{Third}, we collected data on the chosen web applications in their idle state, i.e., when they are open but no user actions are being performed. The purpose of this experiment was to see if application data is transferred when the user does not expect it (an expected data transfer would be because of some action the user takes, e.g. sending an email). For this experiment, we opened each application, logged in if necessary, and then let Wireshark run for 10 minutes while the application sat idle. For simplicity, we only ran these experiments on the Ubuntu/Firefox combination. We chose Ubuntu because we observed less background traffic from Ubuntu compared to Microsoft in the previous experiments. We chose Firefox to try to isolate as much as possible the network activity from the applications themselves: Chrome had the potential for adding additional tracking from the browser, whereas Brave may have blocked trackers that the websites would have otherwise used.

\textbf{Fourth}, we collected data on the web applications listed in the previous section as they would be used in regular everyday usage. The purpose of these experiments was to see if any unexpected endpoints are being contacted during regular usage. We ran Wireshark as we conducted these various activities, and we performed each action on every browser/operating system pair.

The email experiments ran as follows: open the browser, log in, send an email, log out, and close the browser. 

The search experiments ran as follows: open the browser, navigate to the search engine, search for ``Taylor Swift,'' wait for the results to load, and close the browser.

The social networking experiments ran as follows: open the browser, navigate to the site, log in, scroll for one minute, log out, and close the browser. The protocol was slightly different for YouTube: for that experiment we did not log in, we searched for ``Taylor Swift,'' clicked on the first video, and watched for one minute.

Finally, the chat experiments ran as follows: open the browser, navigate to the site, log in, send 3-5 short (between 2 words and one sentence long) messages to the same recipient, and close the browser. (The recipient did not respond during the course of the experiment.)

\subsubsection{Data Analysis}
With our network data, we wanted to determine if the platforms we studied are operating at lower (less data collection, especially unexpectedly) or higher (more data collection) levels of control.

\begin{figure}
    \centering
    \includegraphics[width=\linewidth]{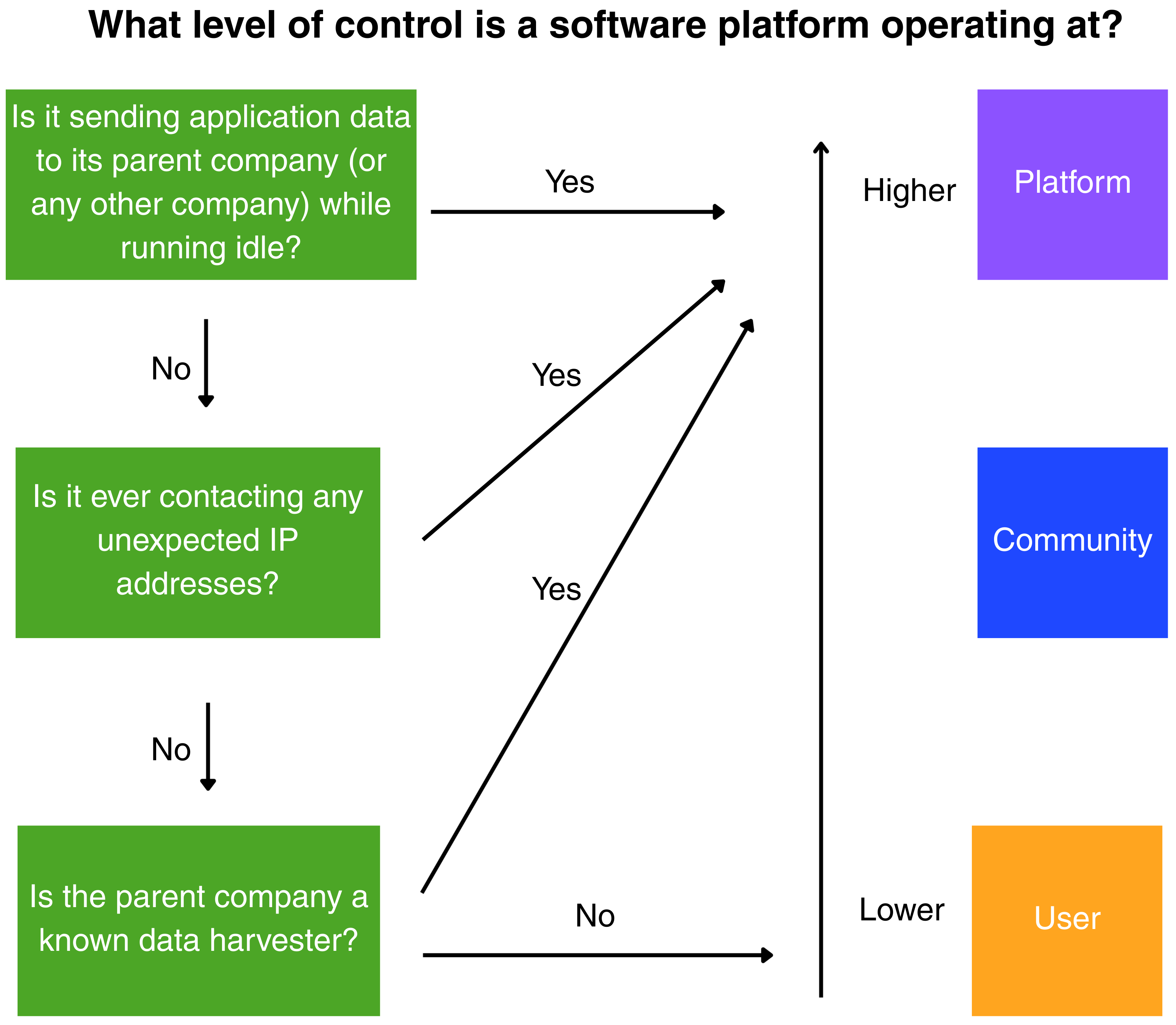}
    \caption{Flow chart to determine which level of control an operating system, browser, or application is operating at. Our data flow monitoring cannot definitively determine if we are operating at a user, community, or platform level of control, but rather determines if we are operating at a lower or higher level of control. A lower level corresponds to operating at closer to the user level of control and a higher level corresponds to operating at closer to the platform level of control, as indicated by the boxes on the right.}
    \Description{Flowchart titled ``What level of control is a software platform operating at?'' On the left side, three stacked green decision boxes guide the flow. On the right side is a vertical upward-pointing arrow indicating the end states of the flow chart. ``Lower'' is written at the bottom and ``higher'' is written at the top, indicating a flow between lower and higher levels of control. The starting green box says, ``Is it sending application data to its parent company (or any other company) while running idle?'' Yes: arrow points directly right toward higher control on the right side arrow. No: arrow points downward to the next decision. ``Is it ever contacting any unexpected IP addresses?'' Yes: arrow points toward higher control. No: arrow points downward to the next decision. ``Is the parent company a known data harvester?'' Yes: arrow points toward higher control. No: arrow points toward lower control. On the other side of the higher/lower arrow, three colored boxes represent levels of control the flow from lowest to highest: Bottom: Orange box labeled ``User. Top: Purple box labeled ``Platform.'' Middle: Blue box labeled ``Community.''
}
    \label{fig:flowchart}
\end{figure}

Through data flow monitoring, because we are observing from the outside, we cannot know if data is only sent to the application for the purpose the user intended (e.g., sending email data to a mail application for the purpose of sending an email, no other data is collected from the user, and the data from the email itself is not used for any other purposes), or if the data sent for an intended purpose is additionally being collected for purposes unintended by the user. Additionally, it is harder to distinguish between community and platform levels of control in data flow monitoring, because that is determined more by \textit{who} the data is being shared with rather than \textit{that} data is shared at all. Thus, we cannot definitively determine if the software operates at user, community, or platform levels of control. Instead, we determine if the software operates at a lower (closer to user-level) or higher (closer to platform or community-level) of control through the following criterion: is application data being sent to the parent company (or any other company) while the software platform is running idle? Additionally, are the platforms only contacting IP addresses affiliated with their parent company and/or known cloud service providers, or are they contacting unexpected IPs (e.g., advertisers or trackers)? Finally, is the parent company a known data harvester (e.g., Google, Meta)? We define the criteria for a ``known data harvester'' as any company for which it is widely known~\cite{schelter2016tracking, gill2013advertising, roesner2012detecting, lerner2016internet, karaj2018whotracks, englehardt2016online, krishnamurthy2009privacy, cassel2022omnicrawl, munir2023cookiegraph} that their business model relies on the harvesting of user data to sell to third parties or to share with third-party advertisers. If the software platforms we examine are doing any of these things we determine that they are operating at a higher level of control; if none, they operate at a lower level of control. The flow chart in Figure \ref{fig:flowchart} outlines our decision-making path for the level of control a software application operates at.

We determined both the IPs being contacted during a particular experiment and the amount of application data being sent to various IPs through the Wireshark logs. We extracted the IP addresses from the Wireshark logs and used the reverse IP address lookup tool \texttt{IPAPI}~\cite{ipapi} to determine the organizations associated with those IP addresses. We focused on the destination IP addresses, that is, the IPs being contacted by the software. We were unable to read the contents of the packets due to encryption, which is a common challenge in work on network monitoring~\cite{hudig2025intimate}.

For the idle OS experiments, we aggregated the number of times each organization was contacted across all of its IPs. We additionally summed the total bytes of application data transferred to that organization's IPs across the hour-long idle time. However, we note that in the TCP protocol, keep alives, heartbeats, and other maintenance-related messages are sent as encrypted ``application data.'' Application data transfers for these purposes are usually small; thus, we only included application data transfers of over 1000 bytes in our summations. While an application data transfer of over 1000 bytes does not guarantee the transfer of user-specific data, it does indicate some significant level of potentially unexpected interaction.

The idle browser experiments were analyzed the same way as the idle OS experiments. However, for an idle browser experiment on a particular OS, we excluded the IPs contacted by that OS in its idle experiment, assuming that contact was from the OS and not from the browser.

The idle application experiments were analyzed the same way as the idle OS and browser experiments. Because these experiments were only conducted on Ubuntu and Firefox, we excluded the background IPs that were collected from the Ubuntu experiment and the Firefox on Ubuntu experiment.

Finally, for the activity experiments, because we would expect application data to be transferred to the parent company while the user is performing an activity, we primarily focus on if the IPs contacted are expected or unexpected.

\begin{table*}[htb]
\caption{\textbf{Participant Demographics}}
\label{tab:participant_demographics}
\begin{tabular}{|cllll|}
\hline
\textbf{ID\#} & \textbf{Gender} & \textbf{Race/Ethnicity} & \textbf{Age} & \textbf{Technical Background}    \\ \hline
P1  &   Male    &   Hispanic        &   30  &   Postdoc researching machine learning, additional expertise in computer security \\ \hline
P2  &   Male    &   White           &   32  &   Undergraduate degree in management with IT concentration   \\ \hline
P3  &   Female  &   White           &   25  &   Everyday technology user    \\ \hline
P4  &   Female  &   Asian     &   29  &   PhD student in computer science researching machine learning \\ \hline
P5  &   Male    &   White           &   23  &   PhD student in mechanical engineering researching robotics and machine learning \\ \hline
P6  &   Male    &   White/Hispanic  &   28  &   PhD student in computer science researching software engineering    \\ \hline
P7  &   Male    &   Hispanic        &   27  &   PhD student in computer science researching machine learning and software engineering \\ \hline
P8  &   Female  &   Asian           &   25  &   Knowledge of basic coding and analytics from undergraduate marketing degree   \\ \hline
P9  &   Male    &   Hispanic        &   25  &   PhD student in computer science researching software engineering  \\ \hline
P10 &   Male    &   White           &   23  &   PhD student in computer science researching machine learning   \\ \hline
P11 &   Female  &   White           &   24  &   Everyday technology user \\ \hline
P12 &   Female  &   White           &   53  &   Everyday technology user \\ \hline
P13 &   Male    &   White           &   57  &   Everyday technology user  \\ \hline
P14 &   Male    &   White           &   42  &   Everyday technology user  \\ \hline
P15 &   Female  &   White           &   31  &   PhD student in computer science researching human-computer interaction \\ \hline
P16 &   Female  &   Black           &   55  &   Everyday technology user    \\ \hline

\end{tabular}
\end{table*}

\subsection{Interviews}

Through data flow monitoring, we determined the level of control that various software platforms operate at. In the next phase of the study, we conducted interviews to understand the level of data control respondents think is \textit{necessary} on these types of software platforms.

\subsubsection{Participant Recruitment}
We recruited a total of 16 participants, at which point little new information was
generated from the interviews, reaching data saturation~\cite{saunders2018saturation}. Participants were required to be everyday computer users and at least 18 years of age. We recruited participants through our university channels and personal networks, and all interview participants were graduate students or employees of our university. Approximately half of participants had a technical background in computer science, human-computer interaction, or mechanical engineering. Two participants had some technical experience from undergraduate business degrees. The other participants were everyday computer users with no specialized knowledge in computer science or a related discipline. Table \ref{tab:participant_demographics} summarizes our participant demographics.

\subsubsection{Interview Protocol}
The interviews were conducted in person or on Zoom (with cameras turned on) by the first author and took a semi-structured format. Rather than asking participants their familiarity with data privacy, which could elicit inconsistent responses on the part of the participants, we instead focused on providing participants with the information about data privacy they needed to make informed responses in the interviews. Participants were first introduced to both the benefits and tradeoffs of data privacy and sharing data with tech companies. The interviewer then shared with the participant the results of our data flow monitoring study to illustrate how common software applications are collecting data, especially in ways that may have been unexpected to the participant. Then, for each of the six software platform types --- operating systems, browsers, email applications, search engines, social media, and chat applications --- participants were led through a thought exercise where they envisioned why it might be necessary for data on those types of applications to be controlled at the platform, community, or user level, as defined in Section \ref{sec:theory}. An example of the thought exercise for search engines is as follows. For platform-level control the interviewer asked, \textit{Pretend that you’re a software engineer at Google working on the search platform. Why might you think that it would be necessary for Google to have control over user data?} For community-level control the interviewer asked, \textit{Pretend that you are the head of a company, university or special interest group that has provided laptops to all its members. For example, pretend you are the head of IT at our university or the principal of a high school that has provided laptops to all its students. Why might it be necessary for you, the head of the organization, to control user data from a search engine?} For user-level control the interviewer asked \textit{Why might it be necessary for the user to have full control over their search data?} A full interview guide is included in the appendix.

The purpose of the thought exercise was to prepare the participants to identify what they think is the necessary level of control for data on that type of software application, and to understand participant motivations for choosing that level of control. The interviewer let the participants come up with their own examples first. The interviewer had her own examples on hand if the participant struggled in order to help the participant make an informed decision. For consistency across participants, even if the participant did not struggle to come up with examples, the interviewer shared all her examples after the participant finished brainstorming. Examples of these for search engines included personalization for a better user experience and advertising to make the product free at the platform level of control, blocking unsafe sites from search results at the community level, and privacy, especially privacy for sensitive data, at the user level. Examples for all platforms are included in the aforementioned interview guide. After the thought exercise, the participant was invited to identify which level of control is simultaneously the lowest level possible and highest level necessary for that type of software application. Participants distinguished between the cases of personal use and group use (encompassing usage in circumstances such as workplaces, schools, and special interest communities) in their responses.

\subsubsection{Data Analysis}
Audio from the interviews was recorded and transcribed. We counted the number of participant responses for platform, community, and user levels of control for each software type in both the personal usage and group usage contexts. We then analyzed the interview transcripts using a reflexive thematic analysis approach~\cite{braun2019reflecting} to better understand participant motivations for selecting each level of control. The first author inductively coded for the reasons participants selected the levels of control that they did for each different software type. The first author also inductively coded for any nuances participants shared in selecting a level of control. Within each of the four categories --- reasons for selecting user level of control, reasons for selecting community level of control, reasons for selecting platform level of control, and nuances in selecting a particular level --- the research team synthesized the codes for that category into overarching themes. Specifically, the first author and second author discussed together the codes the first author identified, discussed and debated how to group the codes into themes, and how to most appropriately name the themes to encapsulate the range of codes they encompass.

\textbf{Positionality Statement:} The research team is entirely Christian, white, and American. The first and second authors who performed the reflexive thematic analysis are female. The advising author, who did not participate in the reflexive thematic analysis process, is male. The first author has a bachelors and masters degree in computer science and is pursuing a PhD in computer science. The second author has a bachelors, masters, and PhD in bioengineering, did postdoctoral work in computer science, and now teaches and researches in computer science and engineering. The third author has a bachelors, masters, and PhD in computer science and now teaches and researches in computer science.

\section{Data Flow Monitoring Results}

In this section we report the results of our data flow monitoring, which gives a sense of if particular platforms are operating at lower or higher levels of control.

We note that although we excluded IP addresses that were confirmed background IPs from a particular operating system or browser, in many cases, previously unseen IP addresses from the same organizations as the background traffic popped up only in the experiments on that particular OS or browser. We report these in the figures and mark them as suspected background interactions. However, we omit these suspected background contacts from the written analysis for clarity and brevity.

\subsection{Idle Operating Systems}

\begin{figure}
    \centering
    \includegraphics[width=\linewidth]{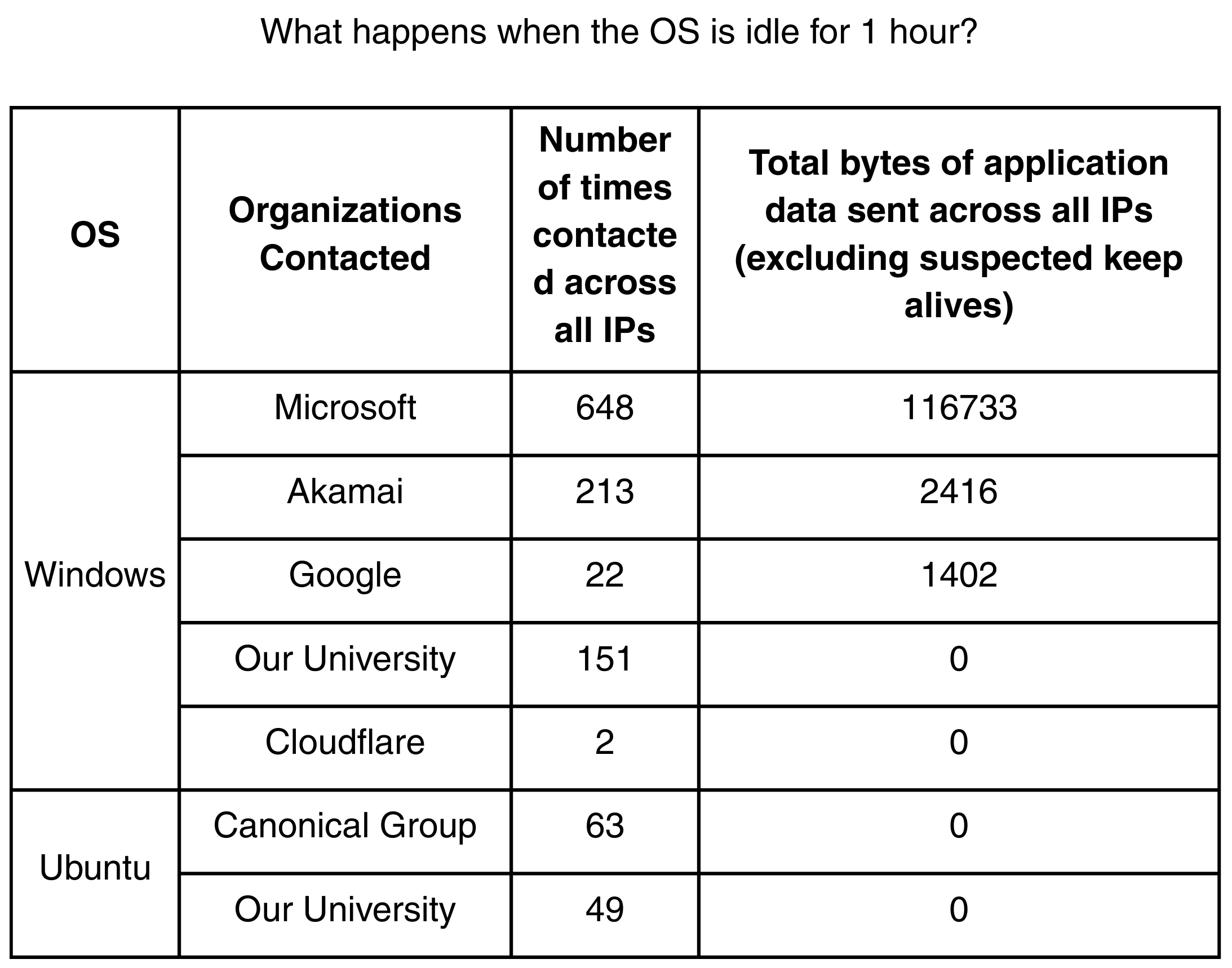}
    \caption{The organizations that the Windows and Ubuntu operating systems contact and transfer application data to when running idle for one hour. In comparison to Ubuntu, Windows contacts more organizations more times. Windows also transfers application data, which is unexpected behavior when running idle. Ubuntu only contacts its parent company and our university (likely because we were on the university Wi-Fi) and does not transfer any application data.}
    \Description{Table titled ``What happens when the OS is idle for 1 hour?'' comparing network activity of Windows and Ubuntu when not in use. Windows: Contacts Microsoft 648 times, sending 152,749 bytes. Contacts Akamai 213 times, sending 3,778 bytes. Contacts Google 22 times, sending 2,262 bytes. Contacts the university 151 times, sending 0 bytes. Contacts Cloudflare 2 times, sending 0 bytes. Ubuntu: Contacts Canonical Group 63 times, sending 0 bytes. Contacts the university 49 times, sending 0 bytes. Key takeaway: Windows shows significantly more idle-time network activity and application data transmission than Ubuntu, which shows only minimal contact and no data sent.}
    \label{fig:idle-os-results}
\end{figure}

In a one hour window, Windows contacted a number of endpoints and transferred application data. Ubuntu contacted Canonical Group (the software company that maintains Ubuntu) and our university, and transferred no application data over 1000 bytes to either. Figure \ref{fig:idle-os-results} summarizes our idle OS results.

\subsection{Idle Browsers}
\begin{figure}
    \centering
    \includegraphics[width=\linewidth]{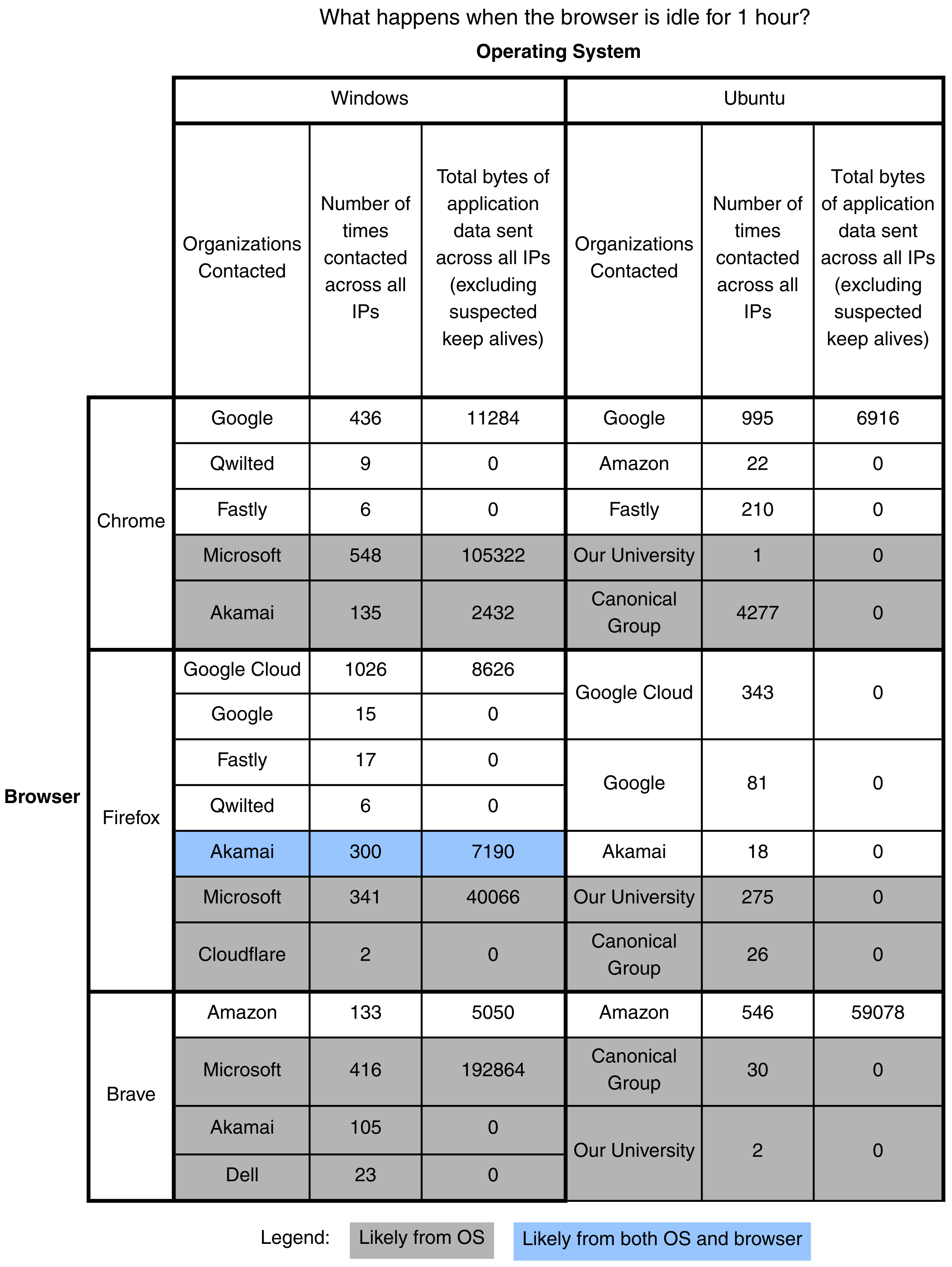}
    \caption{The organizations that Chrome, Firefox, and Brave browsers contact and transfer application data to when they run idle for one hour. We compare the browsers when running idle on Windows and Ubuntu, and highlight organizations for which we suspect contact is background traffic from the operating system rather than traffic from the browser. Contact is suspected to be background traffic from an OS when it is both to an organization contacted in the idle experiment for that OS and does not show up in the idle browser experiments on the other OS. Each browser contacts and transfers application data to its parent company/primary cloud service provider.}
    \Description{Table titled ``What happens when the browser is idle for 1 hour?''  comparing idle-time network activity for Chrome, Firefox, and Brave on Windows and Ubuntu. Legend: Gray shading = Traffic likely originating from the operating system (OS) rather than the browser. Blue shading = Traffic likely originating from both the OS and the browser. No shading = Traffic likely originating only from the browser. Chrome: Windows: Google (436 times, 11,284 bytes), Qwilted (9 times, 0 bytes), Fastly (6 times, 0 bytes), Microsoft (gray) (548 times, 105,322 bytes), Akamai (gray) (135 times, 2,432 bytes). Ubuntu: Google (995 times, 6,916 bytes), Amazon (22 times, 0 bytes), Fastly (210 times, 0 bytes), Our University (gray) (1 time, 0 bytes), Canonical Group (gray) (4,277 times, 0 bytes). Firefox: Windows: Google Cloud (1,026 times, 8,626 bytes), Google (15 times, 0 bytes), Fastly (17 times, 0 bytes), Qwilted (6 times, 0 bytes), Akamai (blue) (300 times, 7,190 bytes), Microsoft (gray) (341 times, 40,066 bytes), Cloudflare (gray) (2 times, 0 bytes). Ubuntu: Google Cloud (343 times, 0 bytes), Google (81 times, 0 bytes), Our University (gray) (275 times, 0 bytes), Canonical Group (gray) (26 times, 0 bytes), Akamai (18 times, 0 bytes). Brave: Windows: Amazon (133 times, 5,050 bytes), Microsoft (gray) (416 times, 192,864 bytes), Akamai (gray) (105 times, 0 bytes), Dell (gray) (23 times, 0 bytes). Ubuntu: Amazon (546 times, 59,078 bytes), Canonical Group (gray) (30 times, 0 bytes), Our University (gray) (2 times, 0 bytes)}
    \label{fig:idle-browser}
\end{figure}

In a one hour window, Chrome transferred application data to Google and contacted content delivery networks. Firefox transferred application data to Google Cloud and contacted content delivery networks. Brave transferred application data to Amazon. Figure \ref{fig:idle-browser} summarizes our idle browser results.

\subsection{Idle Applications}
\subsubsection{Email:} 
\begin{figure}
    \centering
    \includegraphics[width=0.9\linewidth]{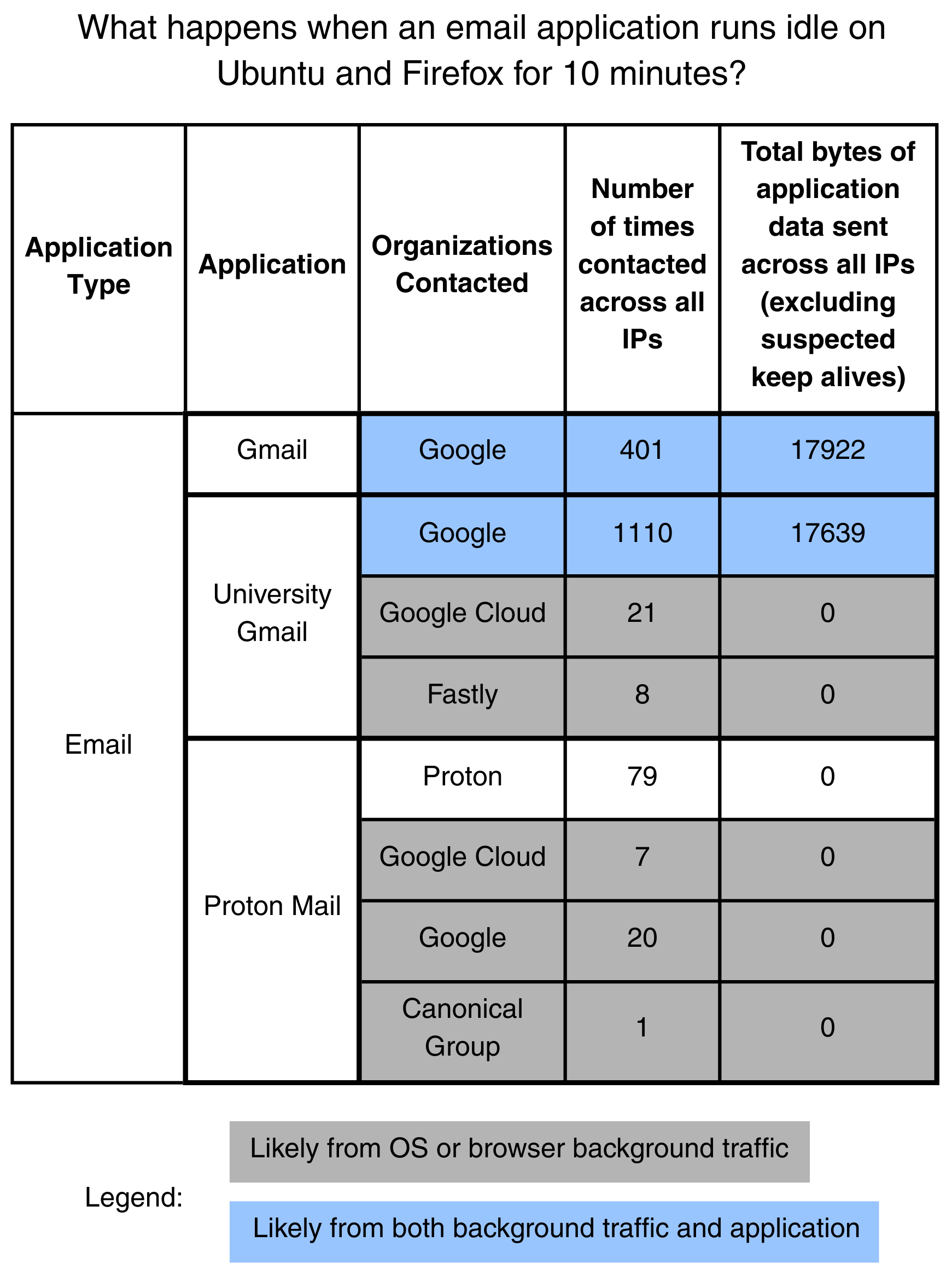}
    \caption{The organizations that various email providers contact and transfer application data to while running idle. Suspected background traffic from the operating system and/or browser is highlighted, as well as cases where traffic could be either background traffic or from the application itself. When both personal and university-sponsored Gmail accounts run idle, they transfer application data back to Google. Proton Mail does not transfer application data while running idle.}
    \label{fig:idle-email}
    \Description{The table, titled ``What happens when an email application runs idle on Ubuntu and Firefox for 10 minutes?,'' examines data traffic for different email applications. The table has five columns: ``Application Type,'' ``Application,'' ``Organizations Contacted,'' ``Number of times contacted across all IPs,'' and ``Total bytes of application data sent across all IPs (excluding suspected keep alives).'' The rows detail the traffic for Gmail, University Gmail, and Proton Mail. A legend indicates gray shading is ''Likely from OS or browser background traffic'' and blue shading is ''Likely from both background traffic and application.'' Gmail: The row for Google is shaded blue, indicating traffic ``Likely from both background traffic and application.'' Google was contacted 401 times, with 17,922 bytes of data sent. University Gmail: The first row, indicating contact with Google, is shaded blue. Google was contacted 1,110 times, with 17,639 bytes of data sent. The other rows for university Gmail are shaded gray, indicating traffic ``Likely from OS or browser background traffic.'' These rows are for Google Cloud (21 contacts, 0 bytes sent) and Fastly (8 contacts, 0 bytes sent). Proton Mail: The first row is for Proton (79 contacts, 0 bytes sent). The other rows are for Google Cloud (7 contacts, 0 bytes sent), Google (20 contacts, 0 bytes sent), and Canonical Group (1 contact, 0 bytes sent), and are all shaded gray, indicating traffic ``Likely from OS or browser background traffic.''}
\end{figure}

Gmail (both personal and university-sponsored) only contacted Google, and transferred application data. Proton Mail contacted Proton and did not transfer any application data. Figure \ref{fig:idle-email} summarizes our idle email results.

\subsubsection{Search:} 
\begin{figure}
    \centering
    \includegraphics[width=\linewidth]{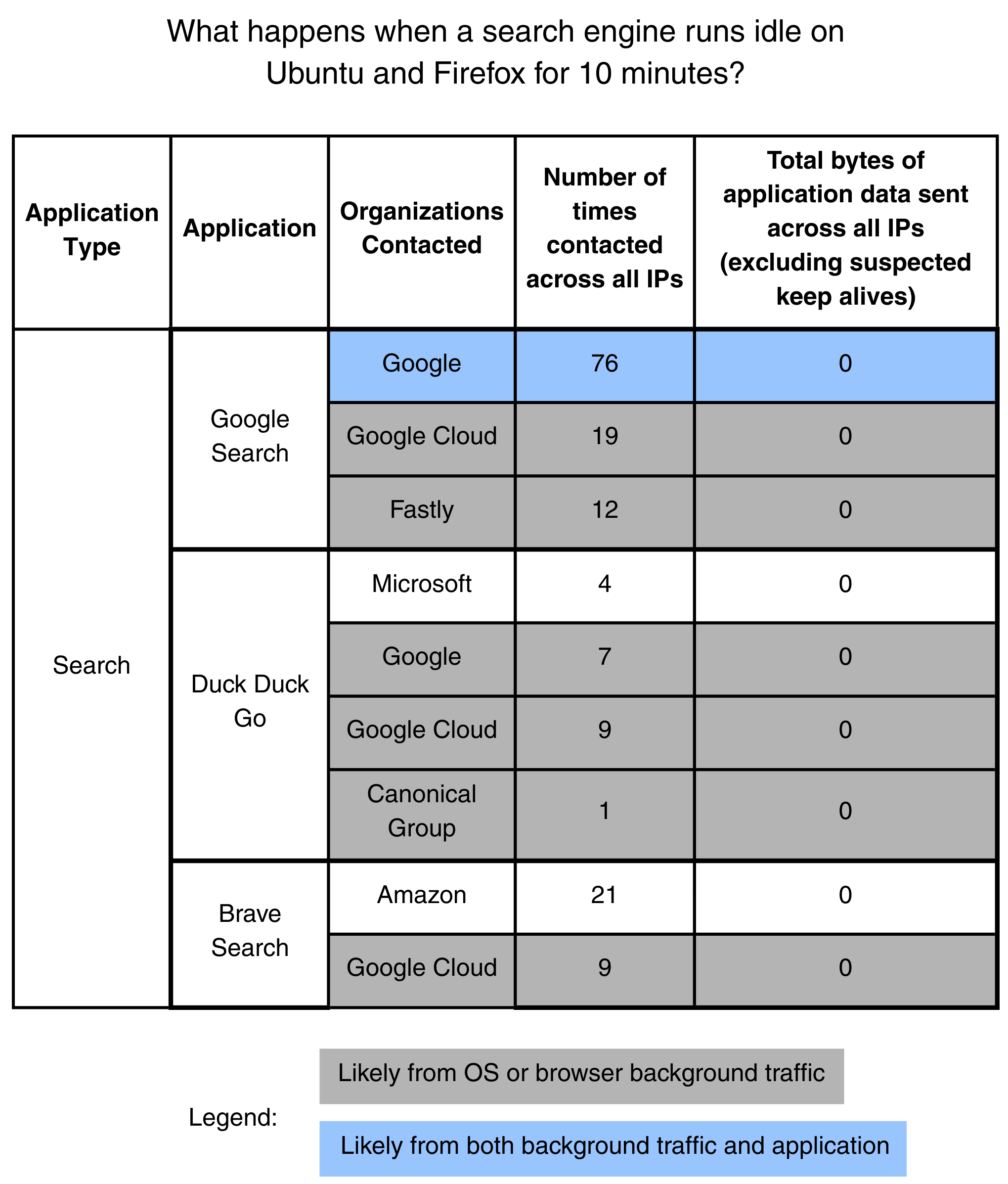}
    \caption{The organizations that various search engines contact and transfer application data to while running idle. Suspected background traffic from the operating system and/or browser is highlighted. While each search engine contacts their parent company or content provider, no search engines transfer application data while running idle.}
    \Description{A table titled ``What happens when a search engine runs idle on Ubuntu and Firefox for 10 minutes?'' The table has five columns: ``Application Type,'' ``Application,'' ``Organizations Contacted,'' ``Number of times contacted across all IPs,'' and ``Total bytes of application data sent across all IPs (excluding suspected keep alives).'' A legend at the bottom indicates that a gray background color represents traffic ``Likely from OS or browser background traffic,'' and a blue background color represents traffic ``Likely from both background traffic and application.'' White shading, though not explicitly in the legend, appears to indicate no significant background traffic. The rows detail the traffic for Google Search, Duck Duck Go, and Brave Search. For Google:  The row for Google is shaded blue, indicating traffic ``Likely from both background traffic and application.'' Google was contacted 76 times, with 0 bytes of data sent. The rows for Google Cloud (19 contacts, 0 bytes sent) and Fastly (12 contacts, 0 bytes sent) are shaded gray, indicating traffic ``Likely from OS or browser background traffic.'' For Duck Duck Go: The row for Microsoft (4 contacts, 0 bytes sent) is shaded white. The rows for Google (7 contacts, 0 bytes sent), Google Cloud (9 contacts, 0 bytes sent), and Canonical Group (1 contact, 0 bytes sent) are shaded gray, indicating traffic ``Likely from OS or browser background traffic.'' For Brave Search: The row for Amazon (21 contacts, 0 bytes sent) is shaded white. The row for Google Cloud (9 contacts, 0 bytes sent) is shaded gray, indicating traffic ``Likely from OS or browser background traffic.''}
    \label{fig:idle-search}
\end{figure}

Google Search contacted Google, Duck Duck Go contacted Microsoft, and Brave Search contacted Amazon. None of these transferred application data. Figure \ref{fig:idle-search} summarizes our idle search engine results.

\subsubsection{Social Media:} 
\begin{figure}
    \centering
    \includegraphics[width=0.82\linewidth]{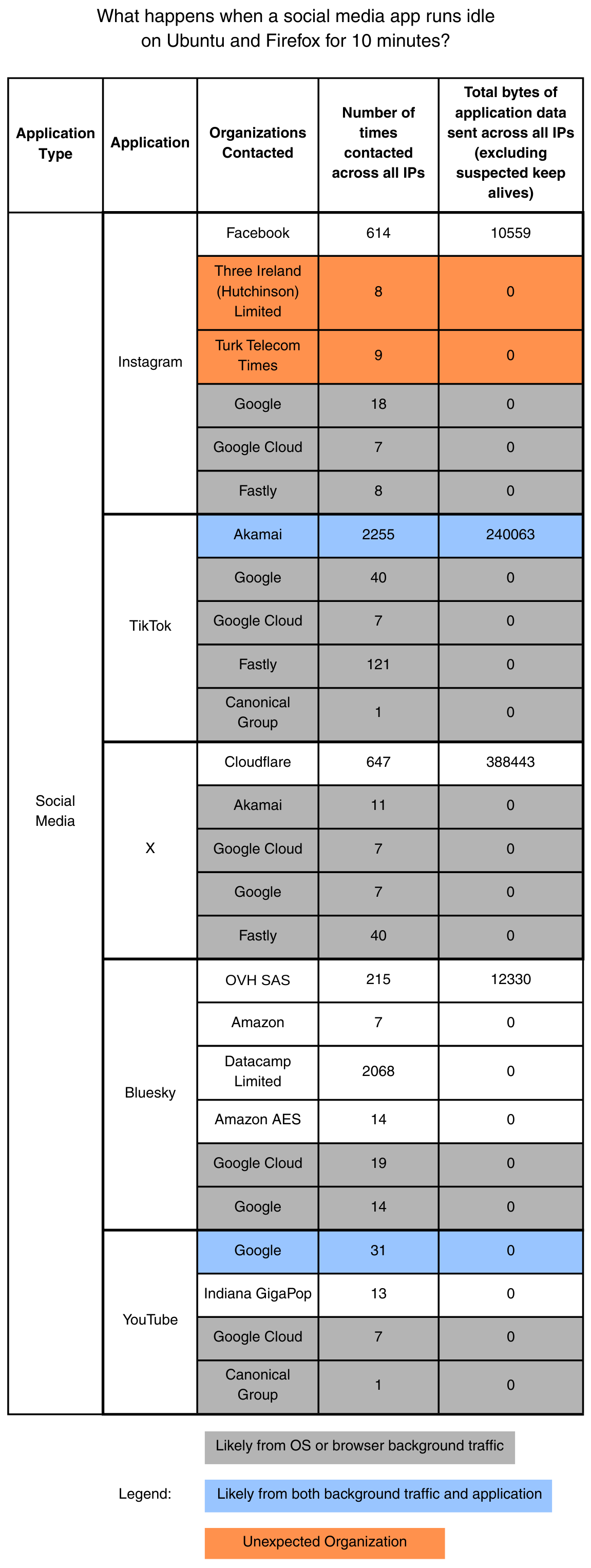}
    \caption{The organizations that various social media applications contact and transfer application data to while running idle. Suspected background traffic from the operating system and/or browser is highlighted. Instagram, Akamai, X, and Bluesky transfer application data back to their parent companies/primary cloud service providers while running idle; YouTube does not. Instagram additionally contacts internet service providers in Ireland and Turkey, which is unexpected.}
    \Description{A table titled ``What happens when a social media app runs idle on Ubuntu and Firefox for 10 minutes?'' The table has five columns: ``Application Type,'' ``Application,'' ``Organizations Contacted,'' ``Number of times contacted across all IPs,'' and ``Total bytes of application data sent across all IPs (excluding suspected keep alives).'' A legend at the bottom indicates three color shadings: Gray: ``Likely from OS or browser background traffic.'' Blue: ``Likely from both background traffic and application.'' Orange: ``Unexpected Organization.'' The rows detail the traffic for Instagram, TikTok, X, Bluesky, and YouTube. For Instagram: The row for Facebook is unshaded. Facebook was contacted 614 times, with 10,559 bytes of data sent. The row for Three Ireland (Hutchinson) Limited is shaded orange, indicating an ``Unexpected Organization.'' It was contacted 8 times, with 0 bytes of data sent. The row for Turk Telekom Times is also shaded orange, indicating an ``Unexpected Organization.'' It was contacted 9 times, with 0 bytes of data sent. The rows for Google (18 contacts, 0 bytes sent), Google Cloud (7 contacts, 0 bytes sent), and Fastly (8 contacts, 0 bytes sent) are all shaded gray, indicating traffic ``Likely from OS or browser background traffic.'' For TikTok: The row for Akamai is shaded blue, indicating traffic ``Likely from both background traffic and application.'' It was contacted 2,255 times, with 240,063 bytes of data sent. The rows for Google (40 contacts, 0 bytes sent), Google Cloud (7 contacts, 0 bytes sent), Fastly (121 contacts, 0 bytes sent), and Canonical Group (1 contact, 0 bytes sent) are all shaded gray, indicating traffic ``Likely from OS or browser background traffic.'' For X: The row for Cloudflare is unshaded. It was contacted 647 times, with 388,443 bytes of data sent. The rows for Akamai (11 contacts, 0 bytes sent), Google Cloud (7 contacts, 0 bytes sent), Google (7 contacts, 0 bytes sent), and Fastly (40 contacts, 0 bytes sent) are all shaded gray, indicating traffic ``Likely from OS or browser background traffic.'' For Bluesky: The rows for OVH SAS (215 contacts, 12330 bytes sent), Amazon (7 contacts, 0 bytes sent), Datacamp Limited (2,068 contacts, 0 bytes sent), and Amazon AES (14 contacts, 0 bytes sent) are unshaded. The rows for Google Cloud (19 contacts, 0 bytes sent), and Google (14 contacts, 0 bytes sent) are shaded gray, indicating traffic ``Likely from OS or browser background traffic.'' For YouTube: The row for Google is shaded blue, indicating traffic ``Likely from both background traffic and application.'' It was contacted 31 times, with 0 bytes of data sent. The row for Indiana GigaPop (13 contacts, 0 bytes sent) is unshaded. The rows for Google Cloud (7 contacts, 0 bytes sent), and Canonical Group (1 contact, 0 bytes sent) are all shaded gray, indicating traffic ``Likely from OS or browser background traffic.''}
    \label{fig:idle-social-media}
\end{figure}

Instagram transferred application data to Facebook and also contacted Three Ireland Limited and Turk Telecom Times. TikTok transferred application data to Akamai: although Akamai is part of Firefox background traffic, the number of contacts to Akamai compared to the other idle experiments, as well as the volume of application data transferred to Akamai, indicate that contact to Akamai is from TikTok as well. X transferred application data to Cloudflare. Bluesky transferred application data to OVH SAS and additionally contacted a few cloud computing and content delivery networks. YouTube contacted Google and Indiana GigaPop and did not transfer any application data while running idle.  Figure \ref{fig:idle-social-media} summarizes our social media results.

\subsubsection{Chat:} 
\begin{figure}
    \centering
    \includegraphics[width=0.8\linewidth]{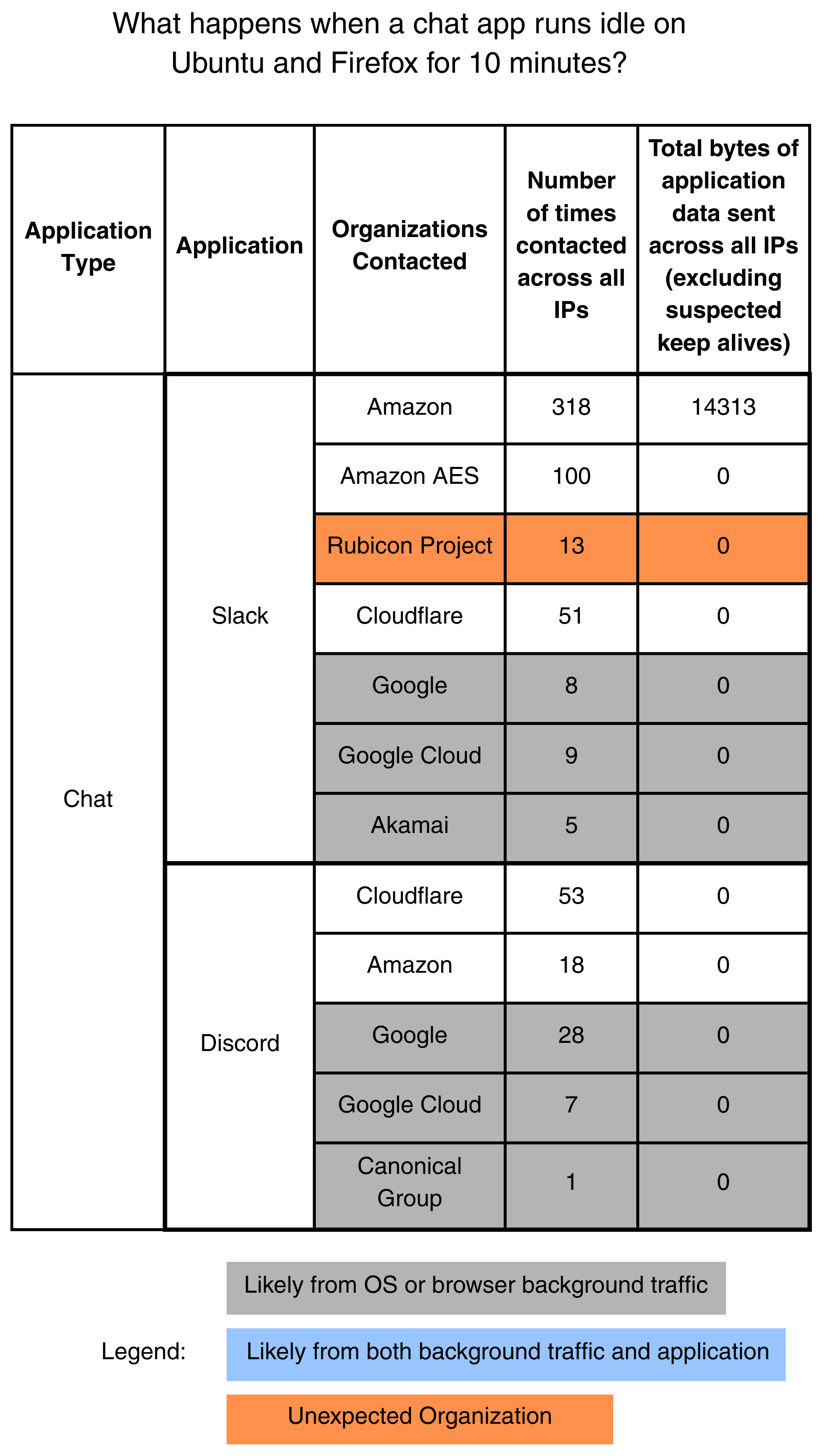}
    \caption{The organizations that various chat applications contact and transfer application data to while running idle. Suspected background traffic from the operating system and/or browser is highlighted. Slack transfers application data to Amazon, its primary cloud service provider, and also pings Rubicon Project, an advertising technology company. Discord does not contact any unexpected organizations and transfers no application data.}
    \Description{A table titled ``What happens when a chat app runs idle on Ubuntu and Firefox for 10 minutes?'' The table has five columns: ``Application Type,'' ``Application,'' ``Organizations Contacted,'' ``Number of times contacted across all IPs,'' and ``Total bytes of application data sent across all IPs (excluding suspected keep alives).'' A legend at the bottom indicates three color shadings: Gray: ``Likely from OS or browser background traffic.'' Blue: ``Likely from both background traffic and application.'' Orange: ``Unexpected Organization.'' The rows detail the traffic for Slack and Discord. For Slack: The row for Amazon (318 contacts, 14,313 bytes sent) is unshaded. The row for Amazon AES (100 contacts, 0 bytes sent) is unshaded. The row for Rubicon Project is shaded orange, indicating an ``Unexpected Organization.'' It was contacted 13 times, with 0 bytes of data sent. The row for Cloudflare (51 contacts, 0 bytes sent) is unshaded. The rows for Google (8 contacts, 0 bytes sent), Google Cloud (9 contacts, 0 bytes sent), and Akamai (5 contacts, 0 bytes sent) are shaded gray, indicating traffic ``Likely from OS or browser background traffic.'' For Discord: The row for Cloudflare (53 contacts, 0 bytes sent) is unshaded. The row for Amazon (18 contacts, 0 bytes sent) is unshaded. The rows for Google (28 contacts, 0 bytes sent), Google Cloud (7 contacts, 0 bytes sent), and Canonical Group (1 contact, 0 bytes sent) are shaded gray, indicating traffic ``Likely from OS or browser background traffic.''}
    \label{fig:idle-chat}
\end{figure}

Slack transferred application data to Amazon and contacted content delivery networks as well as the advertising platform Rubicon Project. Discord contacted Cloudflare and Amazon, and did not transfer any application data. Figure \ref{fig:idle-chat} summarizes our chat application results.

\begin{figure}
    \centering
    \includegraphics[width=\linewidth]{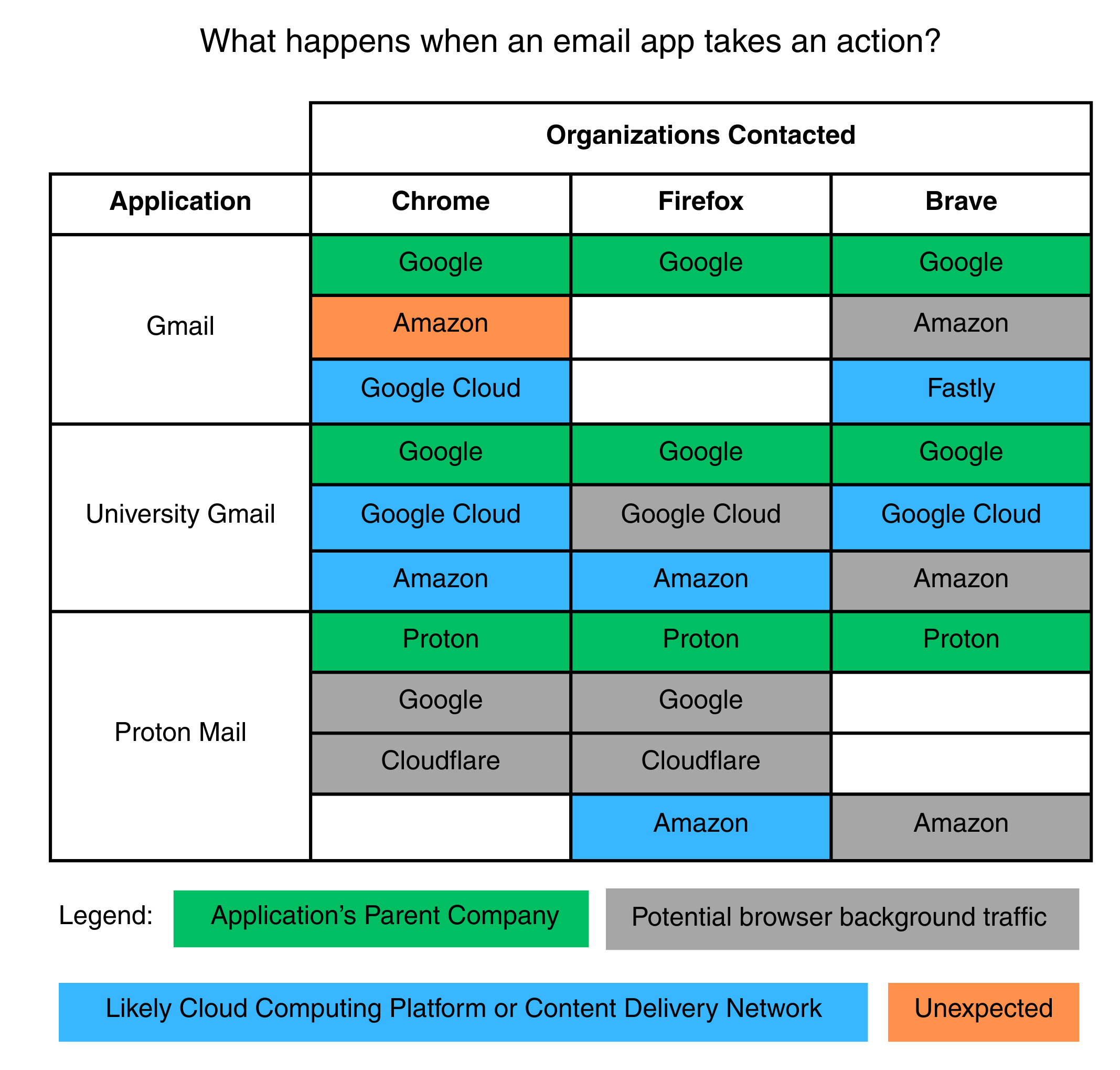}
    \caption{The organizations contacted while sending an email on Gmail, the university-sponsored Gmail, and Proton Mail. While most contacts are expected contacts to the application's parent company or cloud computing/content delivery providers, Gmail's contact of Amazon on Google Chrome is unexpected.}
    \Description{A table titled ``What happens when an email app takes an action?'' The table has four columns: ``Application,'' ``Organizations Contacted'' broken down by Chrome, Firefox, and Brave browsers. A legend at the bottom indicates four color shadings: Green: ``Application's Parent Company.'' Gray: ``Potential browser background traffic.'' Blue: ``Likely Cloud Computing Platform or Content Delivery Network.'' Orange: ``Unexpected.'' The rows detail the traffic for Gmail, University Gmail, and Proton Mail. Gmail: In the Chrome column: Google is green, Amazon is orange, and Google Cloud is blue. In the Firefox column: Google is green. In the Brave column: Google is green, Amazon is gray, and Fastly is blue. University Gmail: In the Chrome column: Google is green, Google Cloud is blue, and Amazon is blue. In the Firefox column: Google is green, Google Cloud is gray, and Amazon is blue. In the Brave column: Google is green, Google Cloud is blue, and Amazon is gray. Proton Mail: In the Chrome column: Proton is green, Google is gray, and Cloudflare is gray. In the Firefox column: Proton is green, Google is gray, Cloudflare is gray, and Amazon is blue. In the Brave column: Proton is green, and Amazon is gray.
}
    \label{fig:action-email}
\end{figure}

\subsection{Application Activity}
For these experiments, we focused only on expected or unexpected contacts rather than amounts of application data transferred. Although we conducted experiments on both Windows and Ubuntu, the results did not differ much other than what we assume to be background traffic. As such, we report only the results from Ubuntu to minimize the background traffic from the OS mixed in with the results. Again, we report the suspected background traffic in the figures but omit it from the written analysis for brevity and clarity.

\subsubsection{Email:}
While sending an email, Gmail contacted Google on all browsers, Amazon on Chrome, Google Cloud on Chrome, and Fastly on Brave. Of these, only Amazon on Chrome is unexpected. The University Gmail contacted Google and Amazon on all browsers. All are expected: Google is the parent company, and Amazon is expected because our university Gmail account uses single sign-on with Okta, which runs on AWS. Proton Mail contacted Proton on all browsers and Amazon on Firefox. The contact with Amazon on Firefox would be unexpected, however, Proton Mail uses AWS for alternative routing~\cite{proton_alternative_routing}, thus, Proton Mail has no unexpected contact. Figure \ref{fig:action-email} summarizes our results.

\subsubsection{Search:} 
\begin{figure}
    \centering
    \includegraphics[width=\linewidth]{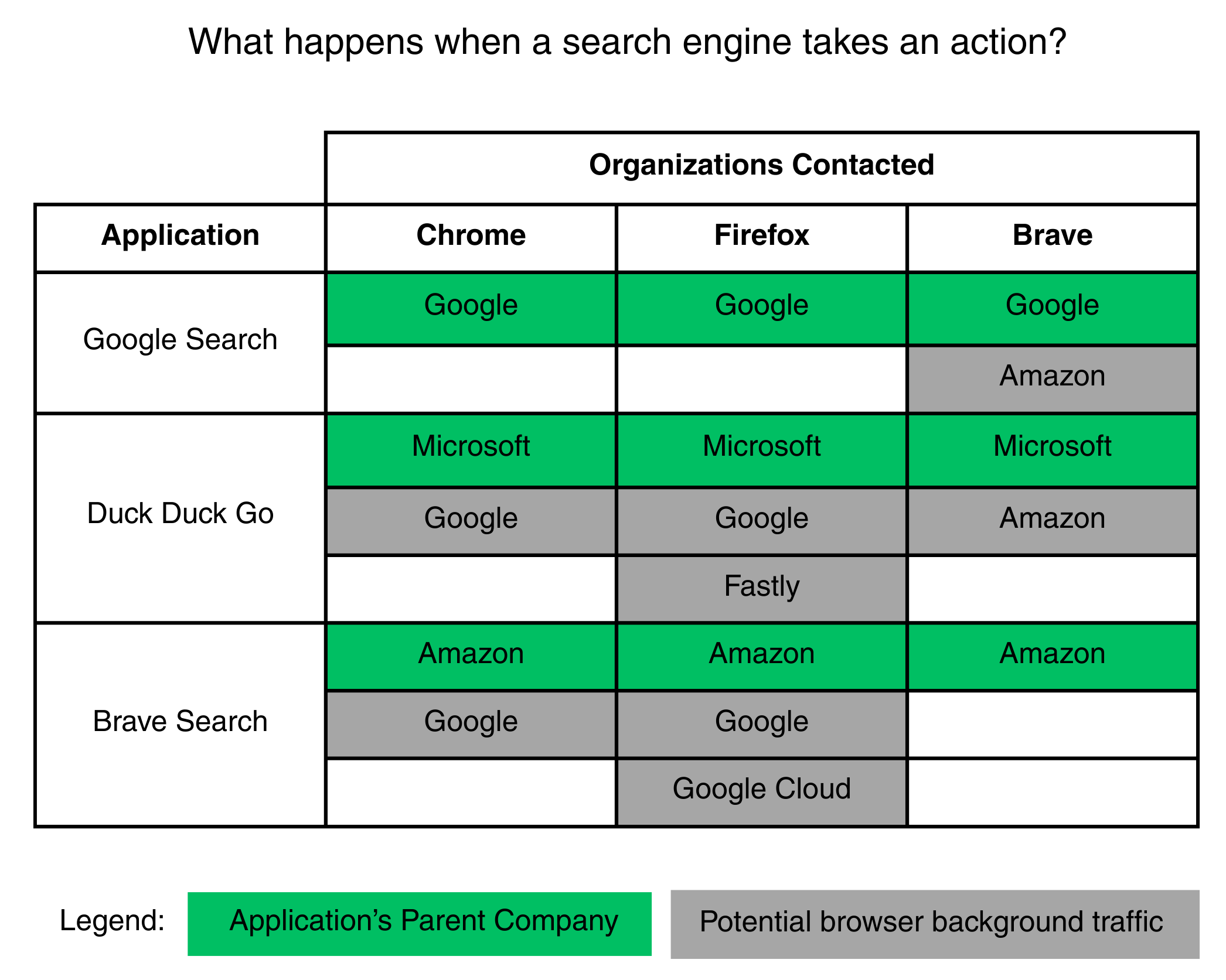}
    \caption{The organizations contacted while performing searches on Google Search, Duck Duck Go, and Brave Search. No contacts are unexpected.}
    \Description{A table titled ``What happens when a search engine takes an action?'' The table has four columns: ``Application,'' ``Organizations Contacted'' broken down by Chrome, Firefox, and Brave browsers. A legend at the bottom indicates two color shadings: Green: ``Application's Parent Company.'' Gray: ``Potential browser background traffic.'' The rows detail the traffic for Google Search, Duck Duck Go, and Brave Search. Google Search: In the Chrome, Firefox, and Brave columns, Google is green, indicating the ``Application's Parent Company.'' In the Brave column, the second row for Amazon is gray, indicating ``Potential browser background traffic.'' Duck Duck Go: In the Chrome, Firefox, and Brave columns, Microsoft is green, indicating the ``Application's Parent Company.'' In the Chrome column, the second row for Google is gray, indicating ``Potential browser background traffic.'' In the Firefox column, the second row for Google is gray, and the third row for Fastly is gray. In the Brave column, the second row for Amazon is gray. Brave Search: In the Chrome, Firefox, and Brave columns, Amazon is green, indicating the ``Application's Parent Company.'' In the Chrome column, the second row for Google is gray, indicating ``Potential browser background traffic.'' In the Firefox column, the second row for Google is gray and the third row for Google Cloud is gray, indicating ``Potential browser background traffic''.}
    \label{fig:action-search}
\end{figure}

Google Search contacted Google on all three browsers. This is expected: Google is the parent company. Duck Duck Go contacted Microsoft on all browsers. This is expected: Duck Duck Go is powered by Microsoft Bing. Brave Search contacted Amazon on all three browsers. This is expected: Brave Search runs on AWS. Figure \ref{fig:action-search} summarizes our results.

\subsubsection{Social Media:} 
\begin{figure}
    \centering
    \includegraphics[width=\linewidth]{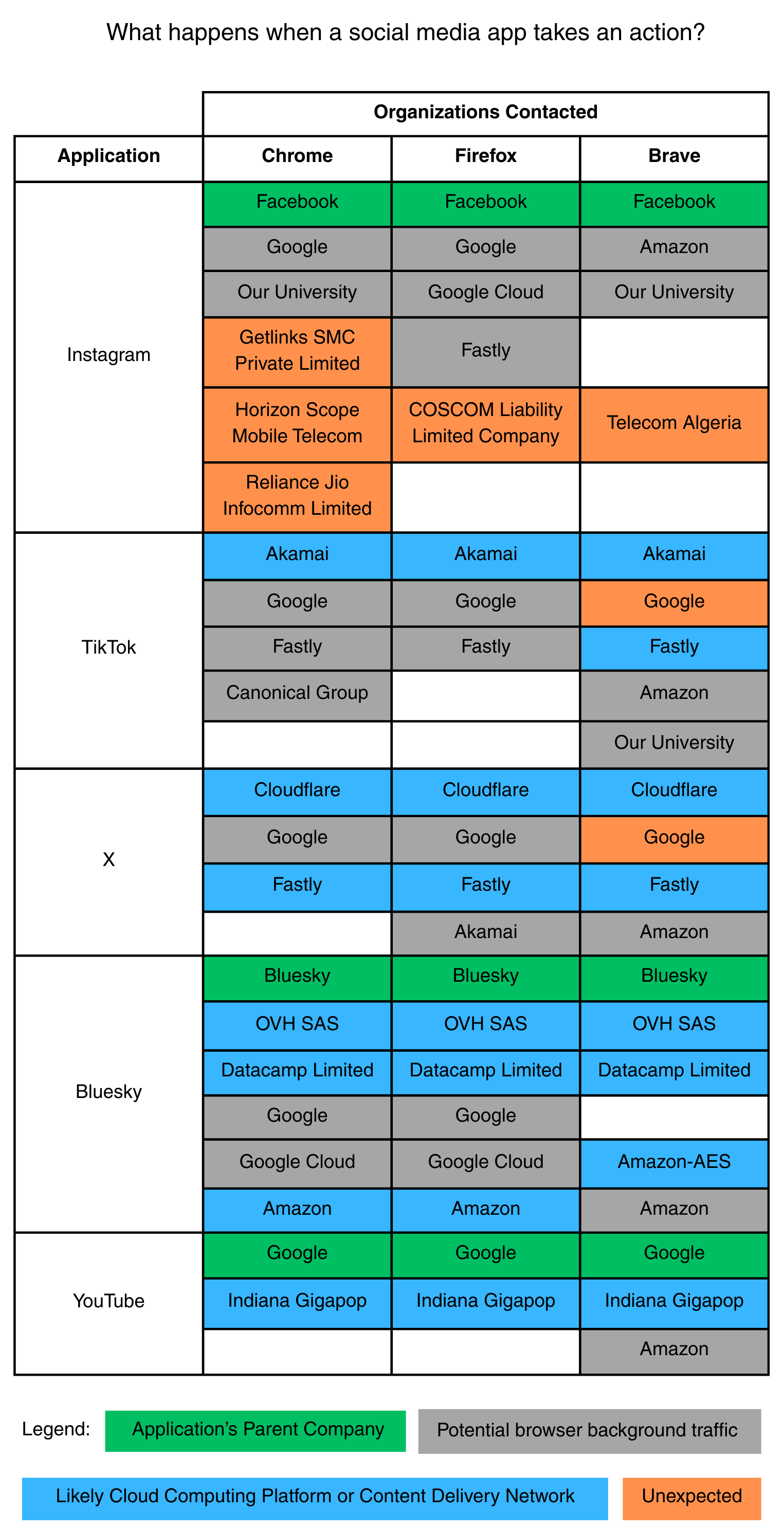}
    \caption{The organizations contacted while using the social media platforms Instagram, TikTok, X, Bluesky, and YouTube. Most contacts were expected; however, Instagram unexpectedly contacted multiple Internet Service Providers in other countries, and TikTok and X unexpectedly contacted Google.}
    \Description{A table titled ``What happens when a social media app takes an action?'' The table has four columns: ``Application,'' and ``Organizations Contacted'' broken down by Chrome, Firefox, and Brave browsers. A legend at the bottom indicates four color shadings: Green: ``Application's Parent Company.'' Gray: ``Potential browser background traffic.'' Blue: ``Likely Cloud Computing Platform or Content Delivery Network.'' Orange: ``Unexpected.'' The rows detail the traffic for Instagram, TikTok, X, Bluesky, and YouTube. For Instagram: In all three browser columns, Facebook is green. In the Chrome column: Google is gray, Our University is gray, Getlinks SMC Private Limited is orange, Horizon Scope Mobile Telecom is orange, and Reliance Jio Infocomm Limited is orange. In the Firefox column: Google is gray, Google Cloud is gray, Fastly is gray, and COSCOM Liability Limited Company is orange. In the Brave column: Amazon is gray and Our University is gray, and Telecom Algeria is orange. For TikTok: In the Chrome and Firefox columns: Akamai is blue, Google is gray, and Fastly is gray. In the Chrome column, Canonical Group is gray. In the Brave column: Akamai is blue, Google is orange, Fastly is blue, Amazon is gray, and Our University is gray. For X: In the Chrome and Firefox columns: Cloudflare is blue, Google is gray, and Fastly is blue. In the Firefox column, Akamai is gray. In the Brave column: Cloudflare is blue, Google is orange, Fastly is blue, and Amazon is gray. For Bluesky: In all three browser columns: Bluesky is green, OVH SAS is blue, and Datacamp Limited is blue. In the Chrome and Firefox columns: Google and Google Cloud are gray, Amazon is blue. In the Brave column, Amazon-AES is blue, Amazon is gray. For YouTube: In all three browser columns: Google is green, and Indiana Gigapop is blue. In the Brave column, Amazon is gray.}
    \label{fig:action-social-media}
\end{figure}

Instagram contacted Facebook on all browsers; this is expected as it is the parent company. Instagram also contacted Internet service providers in other countries: Getlinks SMC Private Limited (Pakistan), Horizon Scope Mobile Telecom (Iraq), Reliance Jio Infocomm Limited (India), COSCOM Liability Limited Company (Uzbekistan), and Telecom Algeria (Algeria). This is unexpected contact given our location in the Midwestern United States. TikTok contacted Akamai, Google, and Fastly on all browsers. Akamai and Fastly are expected due to content delivery; Google is unexpected. X contacted Cloudflare, Google, and Fastly on all browsers. Cloudflare and Fastly are likely content delivery; Google is unexpected. Bluesky contacted Bluesky, OVH SAS, Datacamp Limited, and Amazon on all browsers. This is expected traffic as the parent company and content delivery networks. (Note: Bluesky used to run on Amazon Web Services; now it does not. However given that there could be some lingering contact with AWS due to platform migration, we give Bluesky the benefit of the doubt that its contact with Amazon is not due to unexpected data sharing.) Bluesky additionally contacted Amazon-AES on Brave which is likely for content delivery. YouTube contacted Google and Indiana Gigapop on all browsers. Google is the parent company of YouTube and Indiana Gigapop is a high-speed routing network for academic institutions in the Midwestern United States, so both are expected. Figure \ref{fig:action-social-media} summarizes our results.

\subsubsection{Chat:} Slack contacted Amazon on all browsers, and Amazon-AES, Cloudflare, and Akamai on Chrome and Firefox. Slack runs on AWS so this is expected, and the others are relevant cloud computing or content delivery networks (contacts to Akamai on Firefox could be either content delivery or background traffic). Slack also contacted Google on all three browsers, Fastly on Chrome, and Google Cloud on Chrome and Firefox. These contacts could all be expected background traffic from the browser except for Google on Brave. On Chrome and Firefox, Slack contacted a combination of Microsoft, Rubicon Project, AS-Criteo, RocketFuel, RhythmOne, AS-Outbrain, AppNexus, and Facebook. All of these are unexpected, and all but Microsoft and Facebook are advertising platforms. We note that these organizations are not contacted by Brave, likely due to Brave's ad and tracker-blocking features.

\begin{figure}
    \centering
    \includegraphics[width=\linewidth]{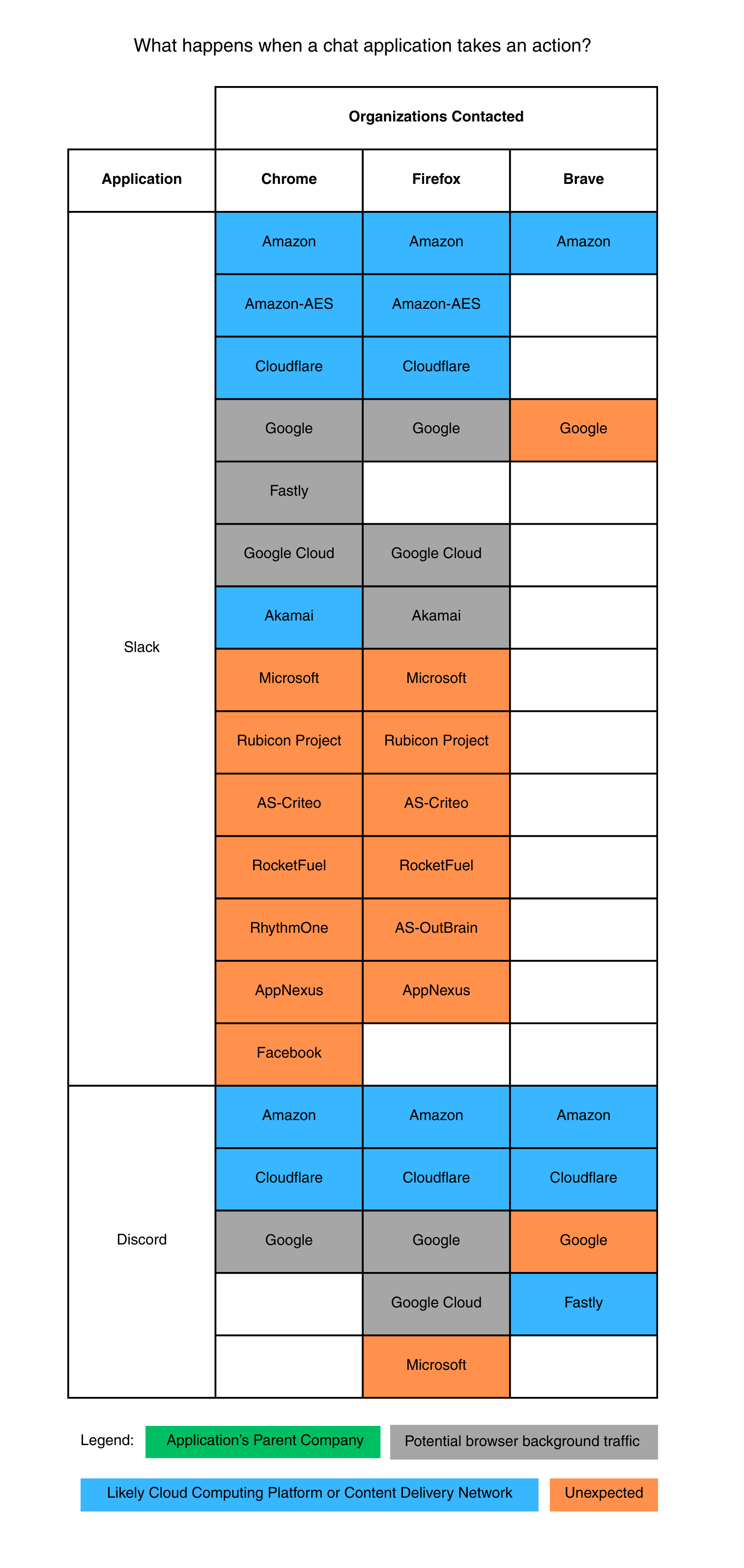}
    \caption{The organizations contacted while sending messages on the chat platforms Slack and Discord. Slack unexpectedly contacted Microsoft, Google, and a number of advertising technology companies. Discord unexpectedly contacted Google and Microsoft.}
    \Description{A table titled ``What happens when a chat application takes an action?'' The table has four columns: ``Application,'' and ``Organizations Contacted'' broken down by Chrome, Firefox, and Brave browsers. A legend at the bottom indicates four color shadings: Green: ``Application's Parent Company.'' Gray: ``Potential browser background traffic.'' Blue: ``Likely Cloud Computing Platform or Content Delivery Network.'' Orange: ``Unexpected.'' The rows detail the traffic for Slack and Discord. For Slack:In all three browser columns, Amazon is blue. In the Chrome and Firefox columns, Amazon-AES and Cloudflare are Blue. Google is listed for all three columns. It is shaded gray for Chrome and Firefox and orange for Brave. In the Chrome column, Fastly is shaded gray. In the Chrome and Firefox columns, Google Cloud is shaded gray. Akamai is listed for both Chrome and Firefox. It is shaded blue for Chrome and gray for Firefox. In the Chrome and Firefox columns, Microsoft, Rubicon Project, AS-Criteo, Rocketfuel, and AppNexus are listed, shared orange. In the Chrome column, RhythmOne and Facebook are listed, shaded orange. In the Firefox column, AS-Outbrain is listed, shaded orange. In the Brave column, Google is listed, shaded orange. For Discord: In all three browser columns, Amazon is blue and Cloudflare is blue. In the Chrome and Firefox columns, Google is gray, and in the Brave column, Google is orange. In the Firefox column, Google cloud is gray and Microsoft is orange. In the Brave column, Fastly is blue.}
    \label{fig:action-chat}
\end{figure}

Discord contacted Amazon and Cloudflare on all three browsers. This is expected: Discord runs on AWS and Cloudflare is a content delivery network. Discord contacted Google on all three browsers, which could be expected background traffic for Chrome and Firefox, but not for Brave. However, we give Discord the benefit of the doubt in its contact with Google due to its ability to integrate with Google products such as Google Drive. Discord additionally contacted Google Cloud on Firefox, expected background traffic. Fastly was contacted on Brave, which is expected as a content delivery network. Microsoft was contacted on Firefox which is unexpected. Figure \ref{fig:action-chat} summarizes our results.

\subsection{Determining Levels of Control}
In this section we determine which level of control a software platform is operating at by considering if application data is transferred while running idle, if any unexpected IPs are contacted, and if the parent company is a known data harvester.

\begin{figure}
    \centering
    \includegraphics[width=\linewidth]{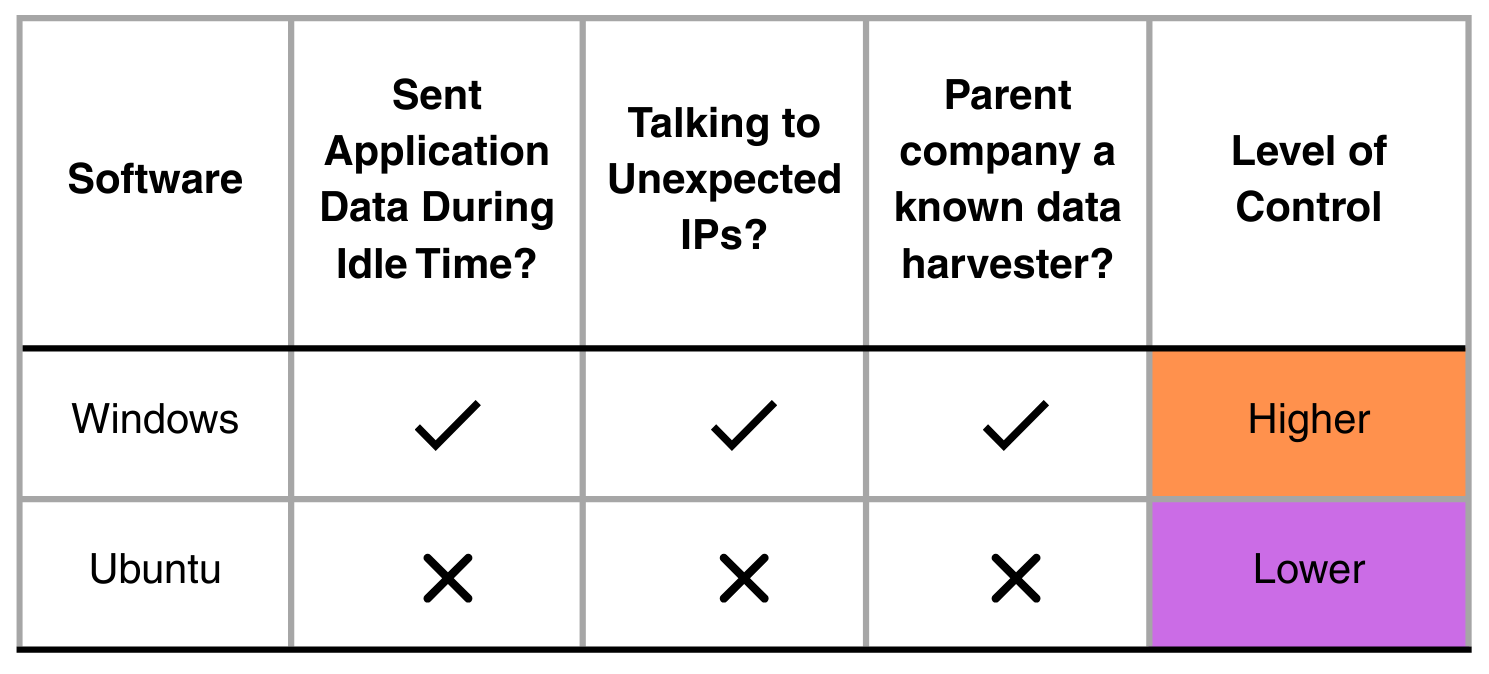}
    \caption{Determining an operating system's level of control based on if it sends application data during idle time, communicates with unexpected IPs, and if its parent company is a known data harvester. If any of these three criteria are met, the operating system operates at a higher level of control; if none are met, it operates at a lower level.}
    \Description{A table with five columns: ``Software,'' ``Sent Application Data During Idle Time?,'' ``Talking to Unexpected IPs?,'' ``Parent company a known data harvester?,'' and ``Level of Control.'' The ``Level of Control'' column is color-coded. The rows detail the findings for Windows and Ubuntu. Windows: ``Sent Application Data During Idle Time?'': A checkmark. ``Talking to Unexpected IPs?'': A checkmark. ``Parent company a known data harvester?'': A checkmark. ``Level of Control'': The cell is shaded orange and contains the text ``Higher.'' Ubuntu: ``Sent Application Data During Idle Time?'': An `x' mark. ``Talking to Unexpected IPs?'': An `x' mark. ``Parent company a known data harvester?'': An 'x' mark. ``Level of Control'': The cell is shaded purple and contains the text ``Lower.''}
    \label{fig:os-levels-of-control}
\end{figure}

\begin{figure}
    \centering
    \includegraphics[width=\linewidth]{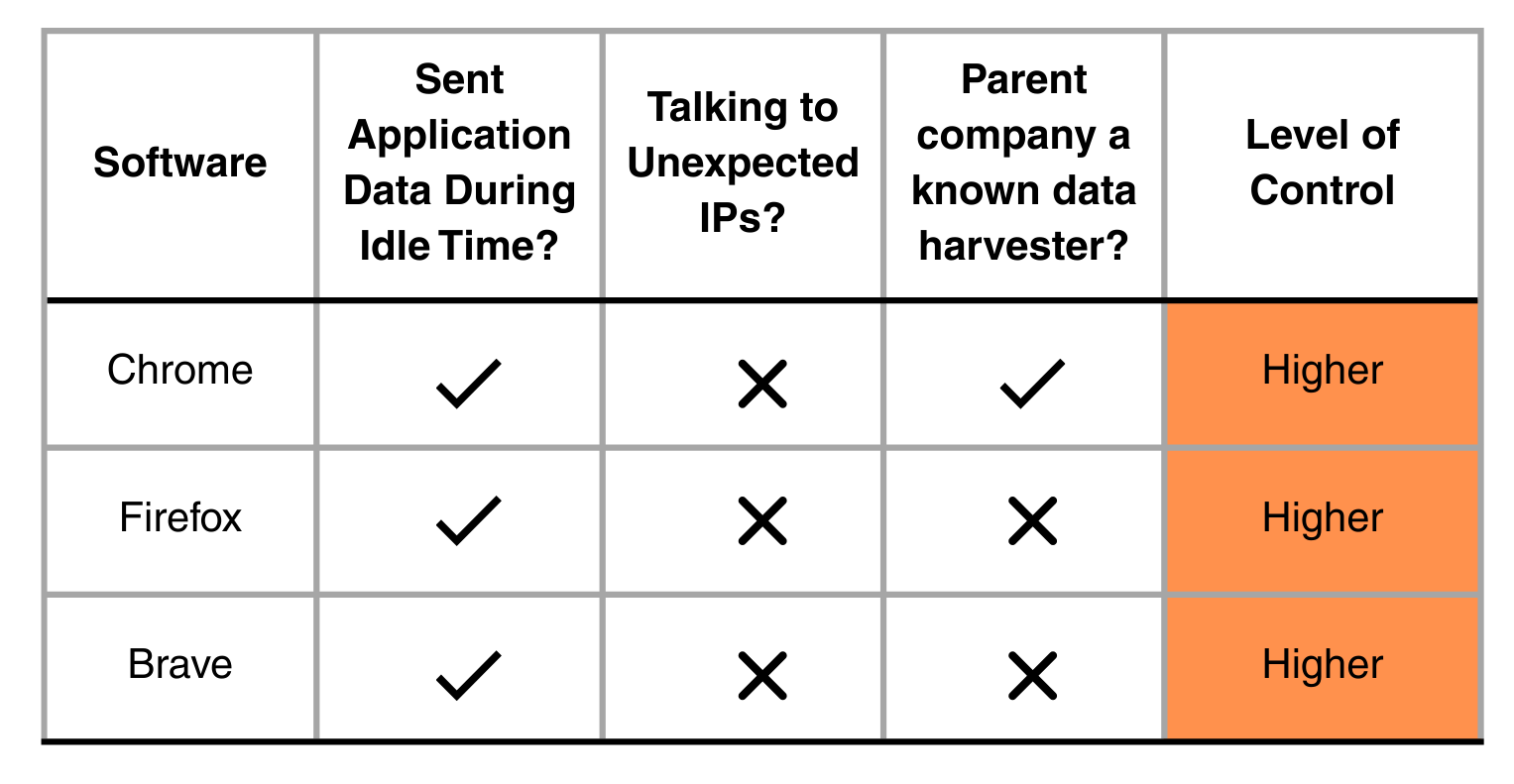}
    \caption{Determining a browser's level of control based on if it sends application data during idle time, communicates with unexpected IPs, and if its parent company is a known data harvester. If any of these three criteria are met, the browser operates at a higher level of control; if none are met, it operates at a lower level.}
    \Description{A table with five columns: ``Software,'' ``Sent Application Data During Idle Time?,'' ``Talking to Unexpected IPs?,'' ``Parent company a known data harvester?,'' and ``Level of Control.'' The ``Level of Control'' column is color-coded. The rows detail the findings for Chrome, Firefox, and Brave. Chrome: ``Sent Application Data During Idle Time?'': A checkmark. ``Talking to Unexpected IPs?'': An `x' mark. ``Parent company a known data harvester?'': A checkmark. ``Level of Control'': The cell is shaded orange and contains the text ``Higher.'' Firefox: ``Sent Application Data During Idle Time?'': A checkmark. ``Talking to Unexpected IPs?'': An `x' mark. ``Parent company a known data harvester?'': An `x' mark. ``Level of Control'': The cell is shaded orange and contains the text ``Higher.'' Brave: ``Sent Application Data During Idle Time?'': A checkmark. ``Talking to Unexpected IPs?'': An `x' mark. ``Parent company a known data harvester?'': An `x' mark. ``Level of Control'': The cell is shaded orange and contains the text ``Higher.''}
    \label{fig:browser-levels-of-control}
\end{figure}

\begin{figure}
    \centering
    \includegraphics[width=\linewidth]{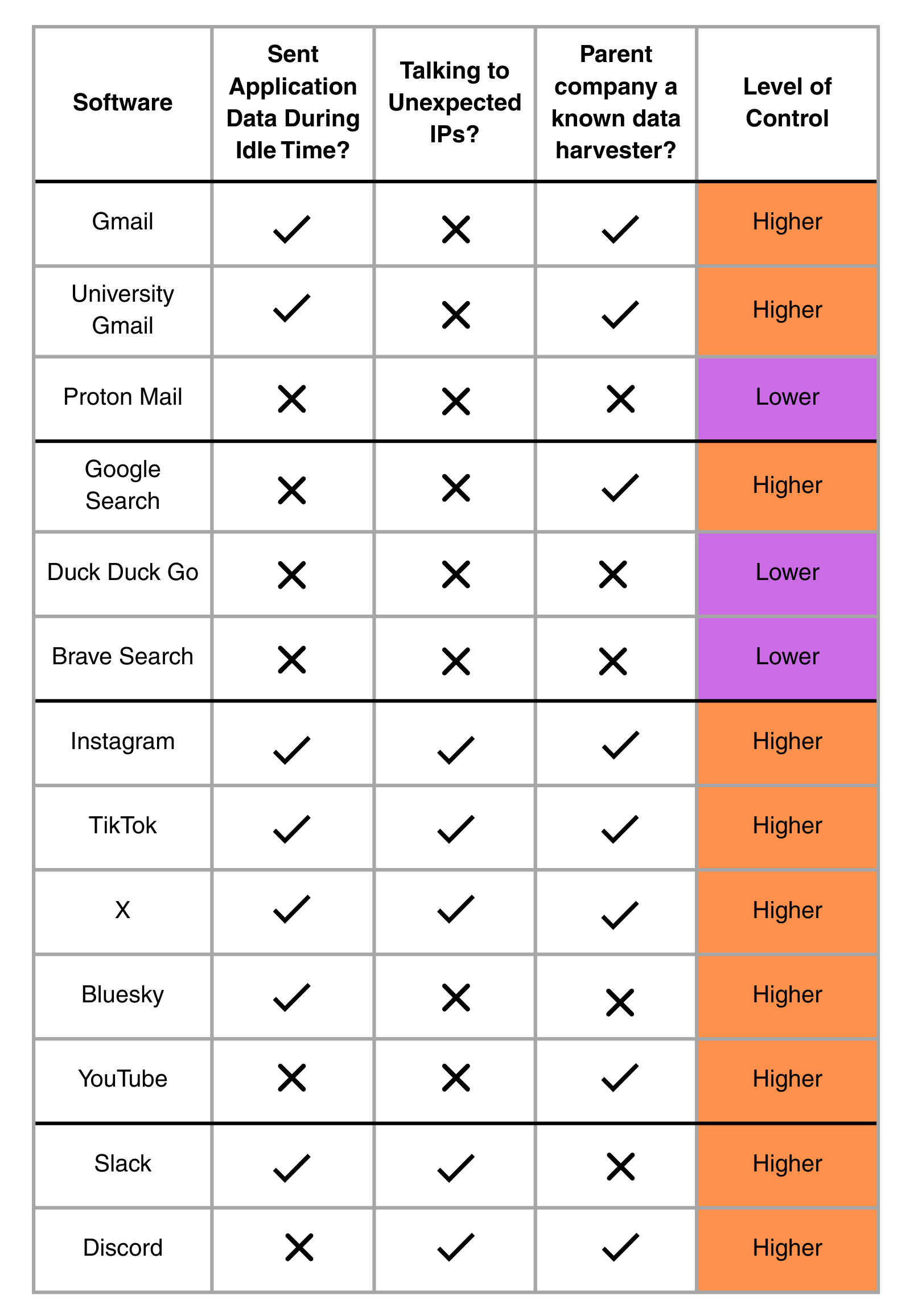}
    \caption{Determining an application's level of control based on if it sends application data during idle time, communicates with unexpected IPs, and if its parent company is a known data harvester. If any of these three criteria are met, the application operates at a higher level of control; if none are met, it operates at a lower level.}
    \Description{A table with five columns: ``Software,'' ``Sent Application Data During Idle Time?,'' ``Talking to Unexpected IPs?,'' ``Parent company a known data harvester?,'' and ``Level of Control.'' The "Level of Control" column is color-coded. The rows detail the findings for various software applications: Gmail: ``Sent Application Data During Idle Time?'': A checkmark. ``Talking to Unexpected IPs?'': An `x' mark. ``Parent company a known data harvester?'': A checkmark. ``Level of Control'': The cell is shaded orange and contains the text ``Higher.'' University Gmail: ``Sent Application Data During Idle Time?'': A checkmark. ``Talking to Unexpected IPs?'': An `x' mark. ``Parent company a known data harvester?'': A checkmark. ``Level of Control'': The cell is shaded orange and contains the text ``Higher.'' Proton Mail: ``Sent Application Data During Idle Time?'': An `x` mark. ``Talking to Unexpected IPs?'': An `x' mark. ``Parent company a known data harvester?'': An `x' mark. ``Level of Control'': The cell is shaded purple and contains the text ``Lower.'' Google Search: ``Sent Application Data During Idle Time?'': An `x' mark. ``Talking to Unexpected IPs?'': An `x' mark. ``Parent company a known data harvester?'': A checkmark. ``Level of Control'': The cell is shaded orange and contains the text ``Higher.'' Duck Duck Go: ``Sent Application Data During Idle Time?'': An `x' mark. ``Talking to Unexpected IPs?'': An `x' mark. ``Parent company a known data harvester?'': An `x' mark. ``Level of Control'': The cell is shaded purple and contains the text ``Lower.'' Brave Search: ``Sent Application Data During Idle Time?'': An `x' mark. ``Talking to Unexpected IPs?'': An `x' mark. ``Parent company a known data harvester?'': An `x' mark. ``Level of Control'': The cell is shaded purple and contains the text ``Lower.'' Instagram: ``Sent Application Data During Idle Time?'': A checkmark. ``Talking to Unexpected IPs?'': A checkmark. ``Parent company a known data harvester?'': A checkmark. ``Level of Control'': The cell is shaded orange and contains the text ``Higher.'' TikTok: ``Sent Application Data During Idle Time?'': A checkmark. ``Talking to Unexpected IPs?'': A checkmark. ``Parent company a known data harvester?'': A checkmark. ``Level of Control'': The cell is shaded orange and contains the text ``Higher.'' X: ``Sent Application Data During Idle Time?'': A checkmark. ``Talking to Unexpected IPs?'': A checkmark. ``Parent company a known data harvester?'': A checkmark. ``Level of Control'': The cell is shaded orange and contains the text ``Higher.'' Bluesky: ``Sent Application Data During Idle Time?'': A checkmark. ``Talking to Unexpected IPs?'': An `x' mark. ``Parent company a known data harvester?'': An `x' mark. ``Level of Control'': The cell is shaded orange and contains the text ``Higher.'' YouTube: ``Sent Application Data During Idle Time?'': An `x' mark. ``Talking to Unexpected IPs?'': An `x' mark. ``Parent company a known data harvester?'': A checkmark. ``Level of Control'': The cell is shaded orange and contains the text ``Higher.'' Slack: ``Sent Application Data During Idle Time?'': A checkmark. ``Talking to Unexpected IPs?'': A checkmark. ``Parent company a known data harvester?;;: An `x' mark. ``Level of Control'': The cell is shaded orange and contains the text ``Higher.'' Discord: ``Sent Application Data During Idle Time?'': An `x' mark. ``Talking to Unexpected IPs?'': A checkmark. ``Parent company a known data harvester?'': A checkmark. ``Level of Control'': The cell is shaded orange and contains the text ``Higher.''}
    \label{fig:app-levels-of-control}
\end{figure}

\subsubsection{Operating Systems:} Windows contacted Google, an unexpected organization, and transferred application data to Microsoft, Akamai, and Google while running idle. Thus, Windows operates at a \textbf{higher} level of control. Ubuntu only contacted Canonical Group (its parent company) and our university (likely related to the Wi-Fi network we were connected to) and transferred no application data to either. Thus, Ubuntu operates at a \textbf{lower} level of control.

\subsubsection{Browsers:} Chrome contacted Amazon which was unexpected, and transferred application data while running idle. Additionally, Chrome is owned by Google: a known data harvester~\cite{schelter2016tracking, gill2013advertising, roesner2012detecting, lerner2016internet, karaj2018whotracks, englehardt2016online, krishnamurthy2009privacy, cassel2022omnicrawl}. Thus, Chrome operates at a \textbf{higher} level of control. 

Although Firefox contacted Google, we believe this is because the Firefox home page contains a Google Search bar. Importantly, no application data was transferred to Google despite the presence of the search bar. However on Windows, application data was transferred to Google Cloud and Akamai. While the transfer of application data to Akamai could have been from the Windows OS, the transfer of application data to Google Cloud suggests that Firefox is operating at a \textbf{higher} level of control.

Brave only contacted Amazon during both its idle experiments, which is expected because Brave is hosted by Amazon Web Services. However, Brave transferred application data to Amazon in both experiments, indicating that Brave operates at a \textbf{higher} level of control.

\subsubsection{Email:}
Gmail and University Gmail transferred application data to Google while running idle. Thus Gmail and University Gmail accounts operate at the \textbf{higher} level of control. Proton Mail did not transfer any application data while running idle, neither did it contact any unexpected organizations while running idle or taking an action. Thus Proton Mail operates at a \textbf{lower} level of control.

\subsubsection{Search:}
None of the search engines transferred any application data while running idle, neither did they contact any unexpected organizations while running idle or taking an action. However, Google Search is owned by Google, a known data harvester. Thus, while Duck Duck Go and Brave Search operate at a \textbf{lower} level of control, Google Search operates at a \textbf{higher} level of control.

\subsubsection{Social Media:}
Instagram, TikTok, and X transferred application data while running idle and contacted a number of unexpected IPs while running idle and/or while taking an action. Bluesky did not contact any unexpected IPs; however, it did transfer application data while running idle. YouTube did not transfer any application data while running idle, neither did it contact any unexpected IPs, however YouTube is owned by Google, a known data harvester. Thus all social media we investigated operate at a \textbf{higher} level of control.

\subsubsection{Chat:}
Slack transferred application data while running idle and contacted unexpected IPs, mostly advertising providers, both while running idle and while taking an action. Discord did not transfer application data while running idle; however, it did unexpectedly contact Microsoft and Google during use. Thus, both applications operate at a \textbf{higher} level of control.

\section{Interview Results}

In this section, we determine what level of control is \textit{necessary} for each software type by reporting the number of participants who identified the user, community, and platform levels of control as necessary in the personal usage and group usage contexts. Then, we report the results of the reflexive thematic analysis of participant motivations for selecting necessary levels of control, which serve as the justifications for moving up to a higher level of control. We also report the common themes from the nuances participants shared in selecting a level of control.

\subsection{What levels of control did participants identify as the necessary level?}

\begin{figure}
    \centering
    \includegraphics[width=\linewidth]{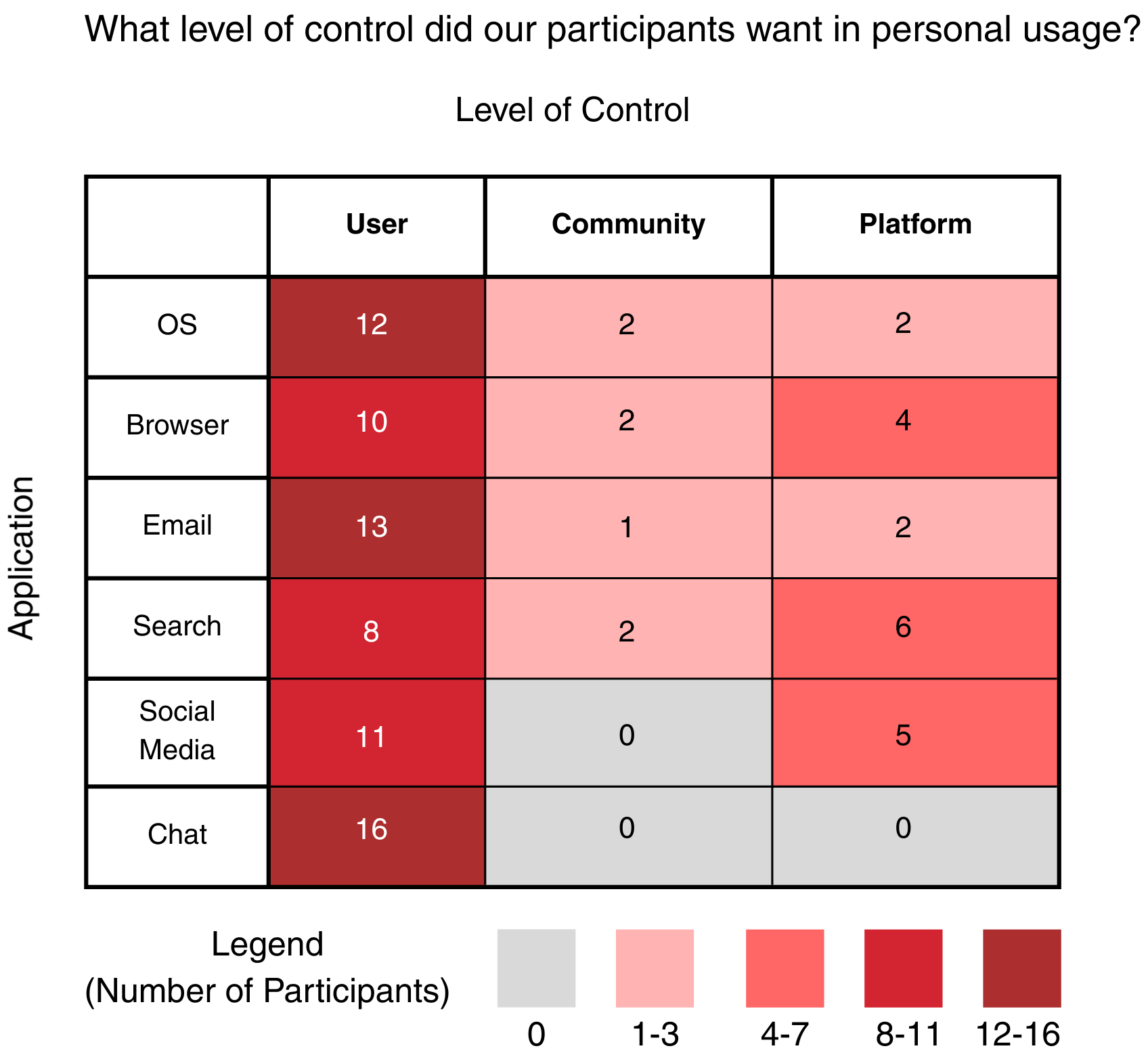}
    \caption{Heat map of participant responses for if they thought user, community, or platform control of their data was necessary in different types of software for personal usage. The number inside each box indicates the number of participants who thought that level of control was necessary for that platform type. The majority of participants wanted user control for all software types. However, more participants wanted platform control for browsers, search engines, and social media.}
    \Description{A heatmap titled "What level of control did our participants want in personal usage?" The table has four columns: ``User,'' ``Community,'' and ``Platform'' under the main heading ``Level of Control,'' and ``Application'' as the row header. The rows are ``OS,'' ``Browser,'' ``Email,'' ``Search,'' ``Social Media,'' and ``Chat.'' A legend on the bottom indicates the color shading based on the number of participants: Gray: 0 participants, Pink: 1-3 participants, Light Red: 4-7 participants, Red: 8-11 participants, Dark Red: 12-16 participants. Here is a breakdown of the data in the heatmap: OS: ``User'' column is dark red with a 12 indicating 12 participants responded with that. ``Community'' column is pink with the number 2. "Platform" column is pink with the number 2. Browser: ``User'' column is red with the number 10. ``Community'' column is pink with the number 2. ``Platform'' column is light red with the number 4. Email: ``User'' column is dark red with the number 13. ``Community'' column is pink with the number 1. ``Platform'' column is pink with the number 2. Search: ``User'' column is red with the number 8. ``Community'' column is pink with the number 2. ``Platform'' column is light red with the number 6. Social Media: ``User'' column is red with the number 11. ``Community'' column is gray with the number 0. ``Platform'' column is light red with the number 5. Chat: ``User'' column is dark red with the number 16. ``Community'' column is gray with the number 0. ``Corporate'' column is gray with the number 0.}
    \label{fig:levels-of-control-personal}
\end{figure}

\begin{figure}
    \centering
    \includegraphics[width=\linewidth]{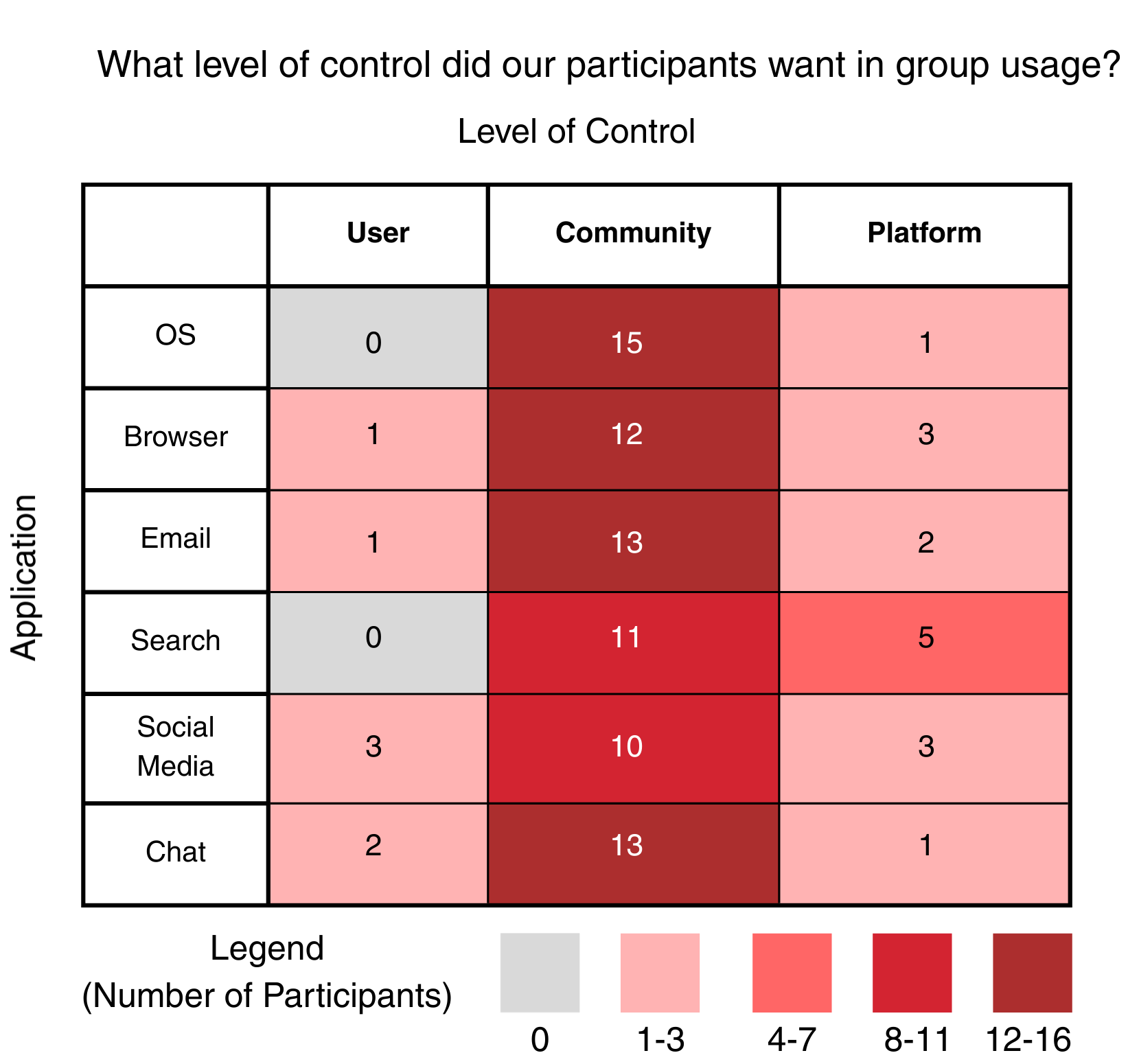}
    \caption{Heat map of participant responses for if they thought user, community, or platform control of their data was necessary in different types of software for group usage. The number inside each box indicates the number of participants who thought that level of control was necessary for that platform type. The majority of participants wanted community control for all software types. However, more participants wanted platform control for search engines.}
    \Description{A heatmap titled ``What level of control did our participants want in group usage?'' The table has four columns: ``User,'' ``Community,'' and ``Platform'' under the main heading ``Level of Control,'' and ``Application'' as the row header. The rows are ``OS,'' ``Browser,'' ``Email,'' ``Search,'' ``Social Media,'' and ``Chat.'' A legend on the bottom indicates the color shading based on the number of participants: Gray: 0 participants, Pink: 1-3 participants, Light Red: 4-7 participants, Red: 8-11 participants, Dark Red: 12-16 participants. Here is a breakdown of the data in the heatmap: OS: ``User'' column is gray with the number 0.``Community'' column is dark red with the number 15. ``Platform'' column is pink with the number 1. Browser: ``User'' column is pink with the number 1. ``Community'' column is dark red with the number 12. ``Platform'' column is pink with the number 3. Email: ``User'' column is pink with the number 1. ``Community'' column is dark red with the number 13. ``Platform'' column is pink with the number 2. Search: ``User'' column is gray with the number 0. ``Community'' column is red with the number 11. ``Platform'' column is light red with the number 5. Social Media: ``User'' column is pink with the number 3. ``Community'' column is red with the number 10. ``Platform'' column is pink with the number 3. Chat: ``User'' column is pink with the number 2. ``Community'' column is dark red with the number 13. ``Platform'' column is pink with the number 1.}
    \label{fig:levels-of-control-community}
\end{figure}

We outline how many participants selected each level of control for the different types of software applications. Results are summarized in the heat maps in Figure~\ref{fig:levels-of-control-personal} for the case of personal usage and Figure~\ref{fig:levels-of-control-community} for the case of group usage.

\subsection{Why did they select these levels of control?}

\subsubsection{User Level:} 
\begin{figure}
    \centering
    \includegraphics[width=\linewidth]{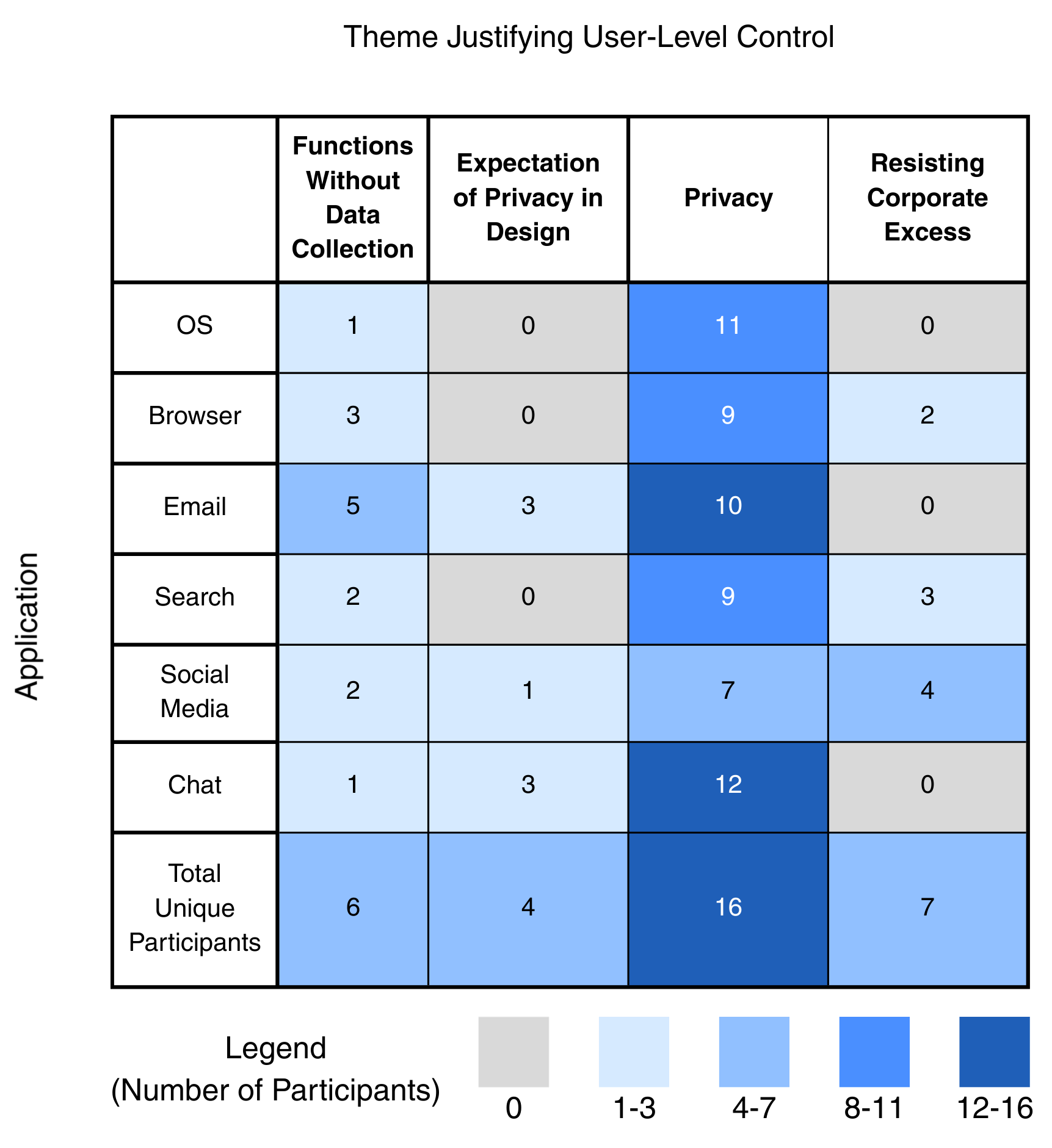}
    \caption{Heat map of themes participants identified to justify user-level control, broken down by each software platform type. We also note the number of unique participants who mentioned this theme across any software platform type. The \textit{privacy} theme was mentioned by the most participants across all software platform types.}
    \Description{A heatmap. The table has five columns: ``Functions Without Data Collection,'' ``Expectation of Privacy in Design,'' ``Privacy,'' and ``Resisting Corporate Excess'' under the main heading ``Theme Justifying User-Level Control,'' and ``Application'' as the row header. The rows are ``OS,'' ``Browser,'' ``Email,'' ``Search,'' ``Social Media,'' ``Chat,'' and ``Total Unique Participants.'' A legend on the bottom indicates the color shading based on the number of participants: Gray: 0 participants, Light Blue: 1-3 participants, Medium-Light Blue: 4-7 participants, Blue: 8-11 participants, Dark Blue: 12-16 participants. Here is a breakdown of the data in the heatmap: OS: ``Functions Without Data Collection'' column is light blue with the number 1. ``Expectation of privacy in design'' is gray with 0. ``Privacy'' is blue with 11. ``Resisting Corporate Excess'' is gray with 0. Browser:  ``Functions Without Data Collection'' column is light blue with the number 3. ``Expectation of privacy in design'' is gray with 0. ``Privacy'' is blue with 9. ``Resisting Corporate Excess'' is light blue with 2.  Email: ``Functions Without Data Collection'' column is medium-light blue with the number 5. ``Expectation of privacy in design'' is light blue with 3. ``Privacy'' is dark blue with 10. ``Resisting Corporate Excess'' is gray with 0. Search:  ``Functions Without Data Collection'' column is light blue with the number 2. ``Expectation of privacy in design'' is gray with 0. ``Privacy'' is blue with 9. ``Resisting Corporate Excess'' is light blue with 3. Social media: ``Functions Without Data Collection'' column is light blue with the number 2. ``Expectation of privacy in design'' is light blue with 1. ``Privacy'' is medium-light blue with 7. ``Resisting Corporate Excess'' is medium-light blue with 4. Chat:  ``Functions Without Data Collection'' column is light blue with the number 1. ``Expectation of privacy in design'' is light blue with 3. ``Privacy'' is dark blue with 12. ``Resisting Corporate Excess'' is medium-light gray with 0. Total unique participants:  ``Functions Without Data Collection'' column is medium-light blue with the number 6. ``Expectation of privacy in design'' is medium-light blue with 4. ``Privacy'' is dark blue with 16. ``Resisting Corporate Excess'' is gray medium-light blue with 7.
    }
    \label{fig:user-themes-heatmap}
\end{figure}

The overarching themes for why participants selected the user level of control were \textbf{functions without data collection}, \textbf{expectation of privacy in design}, \textbf{privacy}, and \textbf{resisting corporate excess.} The number of participants who mentioned each theme per software application type, as well as the total number of unique participants who mentioned each theme across any software application type, can be seen in Figure \ref{fig:user-themes-heatmap}.

\textbf{Functions without data collection} relates to the belief that software applications could still function well without without excessive user data collection that invades privacy, and/or participant desires for a system where they could opt out of features that require excessive data collection in favor of simpler features. As examples, P5 noted that he would be just as happy to use Youtube without autofill and Google Search with its original PageRank algorithm that only relied on website links, not any user data, to rank search results. Similarly, P10 expressed a desire for the ability to opt out of such features that improve user experience but require excessive data collection. P10 also expressed that a simpler advertising model could still generate revenue, appealing to television: \textit{``for TV...you're getting commercials that aren't targeted to you.''} Importantly, some of our participants held that higher-level control was necessary for certain software types but user-level control was possible for others. As an example, while P2 believed that platform-level control was necessary for social media and search to keep them free, he believed that this higher-level control was not necessary for all chat platforms due to different business models employed: \textit{``iMessage is notably, in theory, more secure than a lot of others, because Apple sees it as core to their model of selling iPhones. So the value proposition is coming through something different.''}

\textbf{Expectation of privacy in design} relates to the belief that certain platforms --- in particular, email and chat applications, as well as private social media postings --- create an expectation that data will be private based on the affordances and purpose of the platform. P9 explained, \textit{``I think you get a reasonable expectation that your data isn't getting [collected] because your intent is to talk to a person''} and compared platforms like email and chat to platforms like search engines: \textit{``in one case, you're communicating to...a search service....you're communicating with the machine versus you're communicating between two people when you're communicating via email...If I'm communicating to the machine, whatever I put into the machine that's on me...But in [the case of email or chat], I'm communicating to another person.''} P7 expressed that in the case of social media, when a user makes a post that's private, it creates an expectation that the post will not only be viewable only to their approved connections but also that the data from that post will not be collected by the social media platform. Finally, some participants simply expressed feelings of privacy as why their data should be kept private. P9 said about email, \textit{``I feel like communication is more private''} and P10 said about chat platforms, \textit{``I just feel like it's private when you're messaging the other person.''}

\textbf{Privacy} relates to any user desire to keep their data private, including a general desire for privacy, protecting sensitive data, and freedom of speech and from government surveillance. In relation to operating systems, P6 said, \textit{``Ideally, most of it should be controlled by the user...You could have so much information, like your social security number is probably on there somewhere.''} P7 similarly said, \textit{``I probably have...personal pictures, information, medical records, textual notes...you just want to keep private your own personal information to avoid fraud.''} In the context of email and chat platforms, participants didn't want to \textit{``share any of my information of what I'm talking about with my friends or personal lives''} (P15) and didn't want to have \textit{``sensitive information...leaked''} (P11). Other participants expressed a desire to not have their data be \textit{``sold to somewhere I don't know...or be used for destructive things''} (P11) and also did not want to be tracked. Fear of government surveillance and a desire for the \textit{``freedom to associate''} (P14) was also a concern: \textit{``maybe if someday...the government had ties to certain browsing things...I might feel like I'm not allowed to look certain things up or go to certain news websites...or I might be put on some watch list for going to a certain website''} (P5). Both P13 and P16 mentioned a ``\textit{\textbf{right} to privacy,''} indicating a belief that privacy is more than a mere personal preference.

\textbf{Resisting corporate excess} relates to participant desires to resist corporate overreach in data usage, including the behavioral modification that comes from targeted advertising and recommendation algorithms. As examples, P5 expressed concerns with algorithms \textit{``being able to detect when someone's about to enter a manic episode based off of their behavior on their computer, and they'll start flooding them with ads...and they'll spend a ton of money.''} P14 expressed concerns with algorithms leading to extreme ideologies. P11 expressed concerns with social media's data-powered addictive algorithms impacting real-life relationships: \textit{``say the company is changing your algorithm to get you hooked...you're creating unhealthy habits being online for hours and...you're not creating authentic relationships.''}

\subsubsection{Community Level:}
\begin{figure*}
    \centering
    \includegraphics[width=0.75\linewidth]{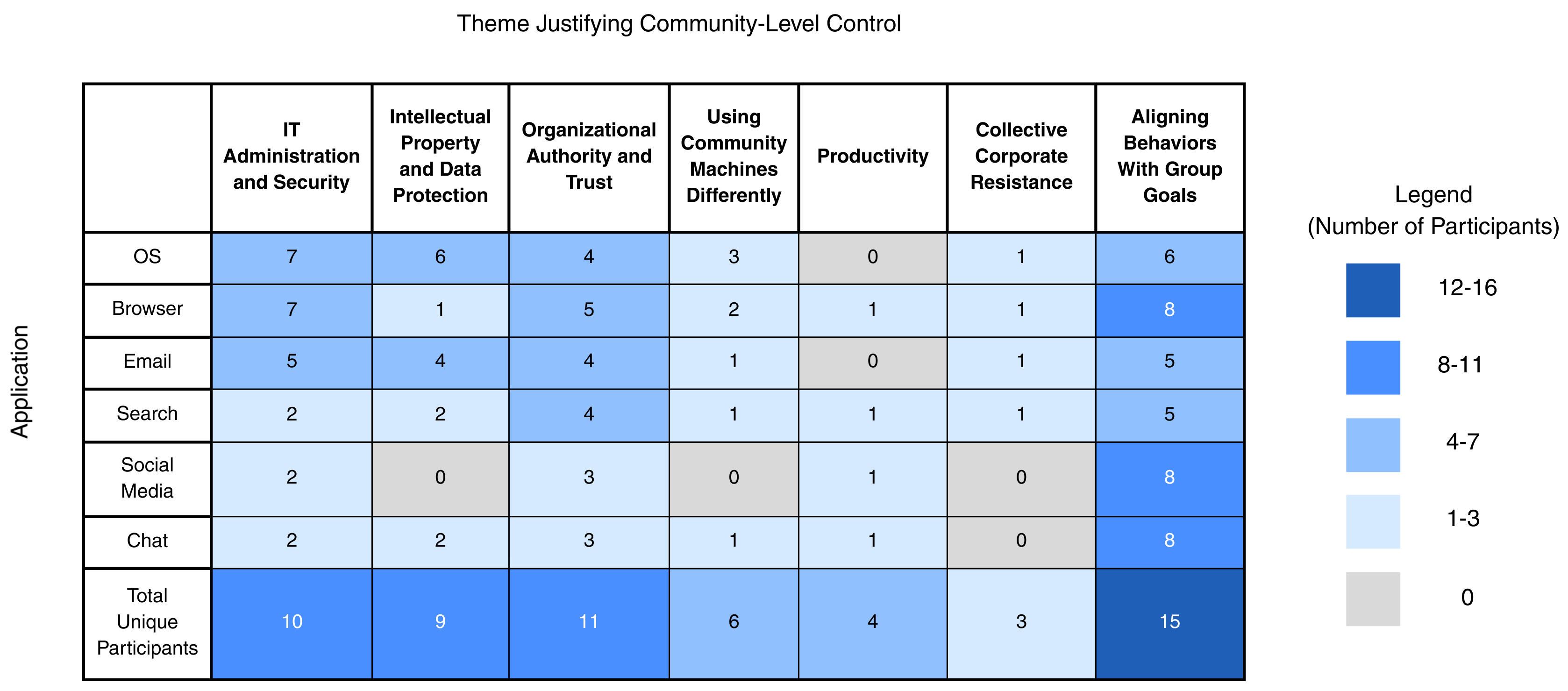}
    \caption{Heat map of themes participants identified to justify community-level control, broken down by each software platform type. We also note the number of unique participants who mentioned this theme across any software platform type. The \textit{aligning behaviors with group goals} theme was mentioned by the most participants across all software platform types. \textit{IT administration and security}, \textit{intellectual property and data protection}, and \textit{organizational authority and trust} also had high levels of responses across our participants.}
    \Description{A heatmap. The table has eight columns: ``IT Administration and Security,'' ``Intellectual Property and Data Protection,'' ``Organizational Authority and Trust,'' ``Using Community Machines Differently,'' ``Productivity,'' ``Collective Corporate Resistance,'' and ``Aligning Behaviors With Group Goals'' under the main heading ``Theme Justifying Community-Level Control,'' and ``Application'' as the row header. The rows are ``OS,'' ``Browser,'' ``Email,'' ``Search,'' ``Social Media,'' ``Chat,'' and ``Total Unique Participants.'' A legend on the right indicates the color shading based on the number of participants: Dark Blue: 12-16 participants, Blue: 8-11 participants, Medium-Light Blue: 4-7 participants, Light Blue: 1-3 participants, Gray: 0 participants. Here is a breakdown of the data in the heatmap: OS: ``IT Administration and Security'' is medium-light blue with the number 7. ``Intellectual Property and Data Protection'' is medium blue with the number 6. ``Organizational Authority and Trust'' is medium blue with the number 4. ``Using Community Machines Differently'' is light blue with the number 3. ``Productivity'' is gray with the number 0. ``Collective Corporate Resistance'' is light blue with the number 1. ``Aligning Behaviors With Group Goals'' is medium-light blue with the number 6. Browser: `IT Administration and Security'' is medium-light blue with the number 7. ``Intellectual Property and Data Protection'' is light blue with the number 1. ``Organizational Authority and Trust'' is medium blue with the number 5. ``Using Community Machines Differently'' is light blue with the number 2. ``Productivity'' is light blue with the number 1. ``Collective Corporate Resistance'' is light blue with the number 1. ``Aligning Behaviors With Group Goals'' is blue with the number 8. Email: `IT Administration and Security'' is medium-light blue with the number 5. ``Intellectual Property and Data Protection'' is medium-light blue with the number 4. ``Organizational Authority and Trust'' is medium-light blue with the number 4. ``Using Community Machines Differently'' is light blue with the number 1. ``Productivity'' is gray with the number 0. ``Collective Corporate Resistance'' is light blue with the number 1. ``Aligning Behaviors With Group Goals'' is medium-light blue with the number 5. Search: `IT Administration and Security'' is light blue with the number 2. ``Intellectual Property and Data Protection'' is light blue with the number 2. ``Organizational Authority and Trust'' is medium-light blue with the number 4. ``Using Community Machines Differently'' is gray with the number 0. ``Productivity'' is light blue with the number 1. ``Collective Corporate Resistance'' is gray with the number 0. ``Aligning Behaviors With Group Goals'' is blue with the number 8. Social media: `IT Administration and Security'' is light blue with the number 2. ``Intellectual Property and Data Protection'' is gray with the number 0. ``Organizational Authority and Trust'' is light blue with the number 3. ``Using Community Machines Differently'' is gray with the number 0. ``Productivity'' is light blue with the number 1. ``Collective Corporate Resistance'' is gray with the number 0. ``Aligning Behaviors With Group Goals'' is blue with the number 8. Chat: `IT Administration and Security'' is light blue with the number 2. ``Intellectual Property and Data Protection'' is light blue with the number 2. ``Organizational Authority and Trust'' is light blue with the number 3. ``Using Community Machines Differently'' is light blue with the number 1. ``Productivity'' is light blue with the number 1. ``Collective Corporate Resistance'' is gray with the number 0. ``Aligning Behaviors With Group Goals'' is blue with the number 8. Total unique participants: `IT Administration and Security'' is blue with the number 10. ``Intellectual Property and Data Protection'' is blue with the number 9. ``Organizational Authority and Trust'' blue with the number 11. ``Using Community Machines Differently'' is medium-light blue with the number 6. ``Productivity'' is medium-light blue with the number 4. ``Collective Corporate Resistance'' is light blue with the number 3. ``Aligning Behaviors With Group Goals'' is dark blue with the number 15.
    }
    \label{fig:community-themes-heatmap}
\end{figure*}

The overarching themes for why participants selected the community level of control were \textbf{IT administration and security}, \textbf{intellectual property and data protection}, \textbf{organizational authority and trust}, \textbf{using community machines differently}, \textbf{productivity}, \textbf{collective corporate resistance}, and \textbf{aligning behaviors with group goals}. We note that while our goal is to assess why a particular level of control is \textit{necessary}, the themes of \textbf{organizational authority and trust} and \textbf{using community machines differently} do not necessarily reflect the \textit{necessity} of community-level control. Rather, these themes help further explain why our participants were willing to surrender control of their data to the community level in exchange for other benefits they found necessary. We report these themes although they are not explanations of necessity because they further illuminate our participant attitudes towards levels of control. The number of participants who mentioned each theme per software application type, as well as the total number of unique participants who mentioned each theme across any software application type, can be seen in figure \ref{fig:community-themes-heatmap}.

\textbf{IT administration and security} pertains to, in a workplace or other community context, allowing system administrators control of machines for purposes such as account management, security, protection against phishing or other online safety, and compliance with software requirements. Our participants found it necessary in professional or other group contexts to have this higher level of control to ensure compliance with group rules and to protect the security of the organization and all members of the group. P6 highlighted the importance of increased security in professional settings saying, \textit{``for a university, there might be more security restrictions, so in that case, they're gonna have a little more control.''} Increased security in community contexts can help protect all members of the community: if one member of the community \textit{``download[s] some malware or virus, [they could] spread the virus across [our university]''} (P7). Community-level control can also help community members receive the technical support they need to use their machines: P6 mentioned the importance of having an IT administrator set up accounts for the virtual machines used in his research lab. Finally, there is a particular importance in having community-level security protections when minors are members of the community. On this note, P12 said, \textit{``working with minors, there's particular sensitivity that they should have protections extended in their digital space....there's lots of things that the quick response from a community-level awareness could could help prevent.''}

\textbf{Intellectual property and data protection} means that in a workplace context, the employer should have control over employee data for the purpose of protecting the company's intellectual property and sensitive data. P11, who works in our university's development office, mentioned the following as a reason why work emails should be controlled by the employer: \textit{``what type of information are we emailing donors? Or the sensitive information that we have about donors: who are we sending that to? It should be only internal.''} Other data to keep private includes \textit{``proprietary information''} (P6) and \textit{``copyright material...[and] intellectual property''} (P14), for example if a company is \textit{``developing some sort of product that's not yet public...and [someone is] emailing it to someone out of the organization''} (P10). P8 and P14 both mentioned keeping data private to protect an organization's image or public relations.

\textbf{Organizational authority and trust} pertains to a willingness to have data be controlled at the community level because of trust in those institutions or belief in their right to control our data as part of our membership. Reasons for trusting such institutions include belief in the mission of the organization, smaller organizations having different incentives than large corporations, and smaller organizations being easier to hold accountable than large corporations. P3 said this about why she trusts her employer, our university, with her data: \textit{``I agree with its values so strongly that it makes me actually have a strong desire to have control on that level, even though I'm sure there are not many people who work for a company that they feel so strongly connected in terms of values.''} Participants also mentioned that our employers have a right to control our work data as part of our employment agreement, and participation in community organizations comes with a degree of buy-in or trust of the community. P2 described this in the professional context: \textit{``the act of entering into a working agreement with a company is very different than just the act of using a service like Google Search. If you're signing up to work at a company or to go to a school, I think there's a much more reasonable degree to which they can have restrictions or have access to certain elements of your data.''} Another element of this theme was a desire from three participants for community collectives to help protect the security of software applications or provide other data-driven features, such that the individual user neither has to do these things themselves nor surrender their data to a large tech company to provide these features.

\textbf{Using community machines differently} encompasses participant comments that they do not use community software for personal usage (especially when that community is their workplace), thus they are comfortable with control at the community level because they are not sharing any personal data. This encompasses both users changing their behavior on community-owned machines (P7 said, \textit{``If it's the work computer, then I shouldn't be putting too much personal information there''}) and users modulating their expectations of privacy on community-owned machines (P15 said, \textit{``I don't necessarily have a problem with them monitoring it if they're not sharing it with anything else, because I'm just using these for professional purposes''}).

\textbf{Productivity} relates to optimizing the use of different tools. In the workplace context this could mean getting \textit{``the sense of the patterns of work, tools that are being used''} (P12) from operating system or browsing data. In the context of a community-run social media platform, this could mean an administrator having access to user data for the purpose of \textit{``tailoring recommendations based on....what the group would be interested in''} (P6). This can help communities run more effectively: P12 noted this necessity in the context of businesses, saying, \textit{``to see what they're paying money for, [if] a program isn't being used....[this] would be helpful so that they can monitor their expenses and their investments and different kinds of programs, and...the general satisfaction from employees with with what they're offering.''}

\textbf{Collective corporate resistance} refers to the sentiment that community organizations could protect users from corporate abuses better than an individual user could protect themselves, thus there is a necessity to have data be controlled at the community level for the purposes of protecting the individual users. P9 said, \textit{``The responsibility shouldn't go to the individual. From an individual level, it's hard to hold the corporation accountable because you don't have power as one person.''}

\textbf{Aligning behaviors with group goals} refers to acceptable use of community-owned machines (e.g., as P5 said, \textit{``if they're installing stuff they're not supposed to be installing, like I gave [a computer] to someone for work and they downloaded Minecraft. Or, if they're mining Bitcoin with their computer instead of using it for school''}), respectful conduct on email, chat, and social media applications (e.g., no hate speech or harassment), and keeping forums on topic (e.g., as P7 said, \textit{``we're a soccer group. We don't want to be publishing about crafts or science''}). Ensuring acceptable use of machines in community contexts can be necessary for the communities to run effectively and achieve their goals.

\subsubsection{Platform Level:}
\begin{figure}
    \centering
    \includegraphics[width=\linewidth]{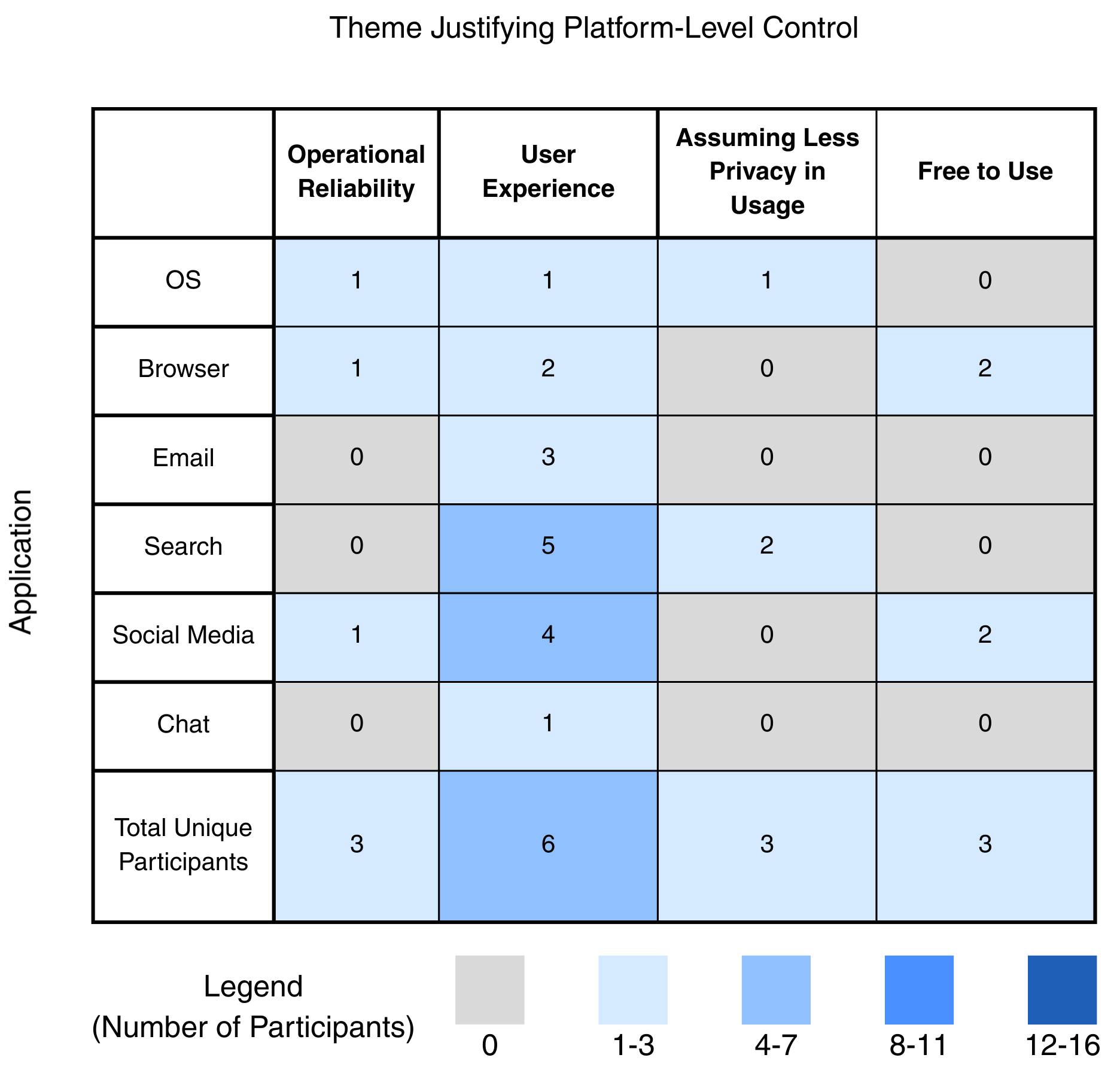}
    \caption{Heat map of themes participants identified to justify platform-level control, broken down by each software platform type. We also note the number of unique participants who mentioned this theme across any software platform type. \textit{User experience} was mentioned by the most participants across all software platform types.}
    \Description{A heatmap. The table has five columns: ``Operational Reliability,'' ``User Experience,'' ``Assuming Less Privacy in Usage,'' and ``Free to Use'' under the main heading ``Theme Justifying Platform-Level Control,'' and ``Application'' as the row header. The rows are ``OS,'' ``Browser,'' ``Email,'' ``Search,'' ``Social Media,'' ``Chat,'' and ``Total Unique Participants.'' A legend on the bottom indicates the color shading based on the number of participants: Gray: 0 participants, Light Blue: 1-3 participants, Medium-Light Blue: 4-7 participants, Blue: 8-11 participants, Dark Blue: 12-16 participants.  Here is a breakdown of the data in the heatmap: OS: ``Operational Reliability'' is light blue with the number 1. ``User Experience'' is light blue with the number 1. ``Assuming Less Privacy in Usage'' is light blue with the number 1. ``Free to Use'' is gray with 0. Browser: ``Operational Reliability'' is light blue with the number 1. ``User Experience'' is light blue with the number 2. ``Assuming Less Privacy in Usage'' is gray with 0. ``Free to Use'' is light blue with 2.  Email:``Operational Reliability'' is gray with 0. ``User Experience'' is light blue with the number 3. ``Assuming Less Privacy in Usage'' is gray with 0. ``Free to Use'' is gray with 0. Search: ``Operational Reliability'' is gray with 0. ``User Experience'' is medium-light blue with the number 5. ``Assuming Less Privacy in Usage'' is light blue with the number 2. ``Free to Use'' is gray with 0. Social media: ``Operational Reliability'' is light blue with the number 1. ``User Experience'' is medium-light blue with the number 4. ``Assuming Less Privacy in Usage'' is gray with 0. ``Free to Use'' is light blue with 2. Chat: ``Operational Reliability'' is gray with 0. ``User Experience'' is light blue with the number 1. ``Assuming Less Privacy in Usage'' is gray with 0. ``Free to Use'' is gray with 0. Total unique participants: ``Operational Reliability'' is light blue with the number 3. ``User Experience'' is medium-light blue with the number 6. ``Assuming Less Privacy in Usage'' is light blue with the number 3. ``Free to Use'' is light blue with 3.}
    \label{fig:platform-themes-heatmap}
\end{figure}

The overarching themes for why participants selected the platform level of control were \textbf{operational reliability}, \textbf{user experience}, \textbf{assuming less privacy in usage}, and \textbf{free to use}. We again note that while our goal is to assess why the platform level of control is \textit{necessary}, \textbf{assuming less privacy in usage} does not necessarily reflect the \textit{necessity} of platform-level control. Again, this theme helps to further explain why our participants were willing to surrender control of their data to the platform level in exchange for other benefits they found necessary. The number of participants who mentioned each theme per software application type, as well as the total number of unique participants who mentioned each theme across any software application type, can be seen in figure \ref{fig:platform-themes-heatmap}.

\textbf{Operational reliability} pertains to a platform's ability to automatically fix bugs and protect security. P3 highlighted the importance of data control at the platform level \textit{``so that my company's IT services are not doing all of the bug crashes and fixes.''} Additionally, P1 expressed concern that without access to data software companies cannot know what needs to be improved: \textit{``if you have no control over any data, no examples...then how can you improve your product security to prevent future attacks?''} (P1).

\textbf{User experience} relates to a platform's ability to improve user experience and build new features because of the large amounts of data collected. Our participants mentioned features like improved AI models and search results, spam filters, content recommendation algorithms, and trend forecasting. The participants who cited user experience as a reason for selecting platform-level control believed that these platforms would not work without access to user data to improve user experience: \textit{``getting information about what that person [is] interested in, recommendations...I think that is necessary for those apps to function and that's why people even get them in the first place''} (P6). One participant even went so far as to say that \textit{``the company...should have a right to see how [the product] is being used''} (P8).

\textbf{Assuming less privacy in usage} means that on certain software platforms or in certain contexts, users will not store sensitive data. Because of the actions they take personally to protect their data, they are comfortable with the platform level of control. P9 explained that he assumes anything he searches on Google could be used as training data for algorithms and thus doesn't search for sensitive things: \textit{``I'm communicating with the machine on my Google searches. So there's kind of a reasonable expectation that that they can train on my data.''} P3, a university employee who works with sensitive student data, shared how she takes personal ownership to protect student data: \textit{``If a student sends me a transcript and it has a social security number on it, I'm gonna delete that file from my computer.''}

\textbf{Free to use} relates to a software application being free to use because of user data collection for targeted advertising and other purposes. Participants who mentioned this theme as a reason for selecting platform-level control found it necessary to have everyday platforms be free: \textit{``I do value having something like a web browser be free''} (P3).

\subsection{What other attitudes emerged from the interviews?}
We asked participants to definitively pick what they thought was the most necessary level of control. In many occasions, the reason why participants chose a particular level was nuanced, or participants wanted to qualify their answers. We coded the nuances and qualifications and grouped these codes into the following themes: \textbf{desiring platform control under better conditions}, \textbf{user-level preference but platform-level necessity}, \textbf{limiting data controlled at the platform level}, and \textbf{limitations of community control}.

\textbf{Desiring platform control under better conditions} embodies the sentiment that our participants would like the benefits of having platforms control their data in an ideal world, but do not trust ``big tech'' to do so responsibly. Some participants expressed a desire for greater regulation on how platforms use their data such that they could trust a platform level of control. P13 described it this way: \textit{``I might think differently if there were stronger safeguards, where people didn't have to worry...I think that government regulation should be the goal.''} All participants who expressed this sentiment ultimately chose a user level of control.

\textbf{User-level preference but platform-level necessity} embodies an opposite sentiment to the previous theme: in this case, participants desired the user level of control of their data but thought the platform level was necessary for its benefits, such as making platforms free. P2 expressed this sentiment as, \textit{``ideally [control] would be at the user level, but I just don't think that's realistic in the current [economic] model.''}

\textbf{Limiting data controlled at the platform level} is the sentiment that while it's important to have some data be controlled at the platform level, this data should be limited in scope, for example data about operating system crashes, or should require informed consent before sharing. P12 said, \textit{``some of the data I don't mind [sharing]...the system operations, technical stuff.''}

\textbf{Limitations of community control} encompasses participant concerns with the community level of control even when selecting that as the necessary level. Some participants particularly highlighted concerns with worker surveillance when the community in question is an employer. P13 said, ``I would be skeptical of an employer wanting to monitor by the minute employee productivity so that they can figure out ways to extract more labor that are potentially dehumanizing.''

\section{Comparison of Actual Levels of Control and User-Identified Necessary Levels of Control}

In this section, we will discuss if subsidiarity is upheld for the different software application types we studied, comparing our data flow monitoring results to the necessary level of control identified in the interviews. Although our data flow monitoring was only able to distinguish ``lower'' and ``higher'' levels of control, for comparison we will equate ``lower'' control with user-level control, and ``higher'' control with platform-level control. We only consider interview participant responses for necessary level of control in the personal usage case, since we did not detect data flows to community organizations. (We will address community-level control in the discussion.) Although a majority of our interview participants thought user-level control was necessary for all software types, we distinguish between cases of a \textit{strong majority} (12-16 participants selecting user-level control) and a \textit{weak majority} (8-11 selecting user-level control). In the cases of weak majority, we discuss options for both user and platform level of control.

For \textbf{operating systems}, a strong majority wanted user-level control. We found that Ubuntu provides lower-level control, so subsidiarity is able to be achieved for operating systems.

For \textbf{browsers}, a weak majority wanted user-level control. None of the browsers we studied provided lower-level control. For the weak minority wanting platform-level control of browsers, subsidiarity is able to be achieved with the browsers we studied.

For \textbf{email}, a strong majority wanted user-level control. We found that Proton Mail provides lower-level control, so subsidiarity is able to be achieved for email.

For \textbf{search engines}, a weak majority wanted user-level control. We found that both Duck Duck Go and Brave Search provide lower-level control. The weak minority wanting platform-level control can use Google Search, which we found operates at a higher level of control. Thus, subsidiarity is able to be achieved for search engines.

For \textbf{social media}, a weak majority wanted user-level control. None of the social media applications we studied operate at a lower level of control. Thus, subsidiarity is not achieved with those social media applications.

For \textbf{chat applications}, a strong majority wanted user-level control. None of the chat applications we studied operate at a lower level of control. Thus, subsidiarity is not achieved with those chat applications.

\section{Discussion}

We discuss the design implications of our findings, including how the principle of subsidiarity can inform what to build, the technical and logistical challenges of implementing user-level control, the implications of greater trust in communities over corporations, limitations, and opportunities for future work.

\subsection{Design Implications: Violations of Subsidiarity Can Show Us What To Build}
Given that willingness to share data can be contextual~\cite{hudig2025intimate} and that some users find data-sharing to be useful~\cite{ur2012smart}, the principle of subsidiarity helps us discern when control should be in the hands of the user and when it is necessary to move control to a higher level to achieve some better outcome for users. In light of this, we will discuss two software types with particularly salient findings from this study: chat applications and search engines.

\subsubsection{Improving Chat Applications}
Chat applications were the only software type where our participants unanimously found user-level control necessary for personal usage. Yet, neither of the chat applications we considered offered user-level control, and Slack was the most egregious across all software types in contacting advertisers. Furthermore, one of the themes for why participants found privacy necessary in chat applications was because of an \textit{expectation of privacy because of the design and purpose of the application}. In other words, because chat applications make the user think they are having a private conversation, there is a greater expectation of privacy in comparison to applications like search engines. While some chat platforms not included in our study may offer user-level control (for example Telegram, especially with its strictest privacy controls), our interview results indicate that it is unacceptable that \textit{any} chat application would not have strong protections over user privacy. Especially given that a majority of users have negative feelings towards data tracking but don't take any steps to minimize tracking themselves~\cite{coopamootoo2022feel}, chat applications should be better designed to protect user privacy by default. This could be achieved by following design guidelines such as \textit{Privacy by Design}~\cite{cavoukian2009privacy} or requiring informed consent. In light of prior findings that users make different choices with their data when offered greater transparency~\cite{liccardi2014no, van2017better}, a platform's interface design can help achieve such transparency. Platforms designed in ways that make users think they are private (such as chat interfaces) should not collect data unexpectedly, and conversely, platforms that will collect data should be designed in ways that give the user this expectation.

\subsubsection{Search}
With search engines, we found a situation almost opposite to that with chats. In the case of search engines, more of our participants were open to platform-level control. However, two out of the three search engines that we evaluated operated at user-level control. While it is still important for these platforms to exist to cater to those who want user-level control, there is a discrepancy: more users find platform-level control necessary for search engines compared to chat applications, yet there are multiple options for search engines that offer user-level control. Considering which software types have the most votes for user-level control can help us prioritize which software types to build more privacy-focused alternatives for.

\subsection{Technical and Logistical Challenges of Implementing User-Level Control}
Even with available options at desired levels of control, technical or logistical challenges may exist with employment, particularly at the user level. For example, someone may desire an operating system with user-level control but not have the technical ability to install an OS like Linux. Similarly, someone may prefer an email service with user-level control but be forced to use a platform like Gmail by their employer. Thus, we cannot achieve subsidiarity only by having a myriad of options to choose from: other systemic changes are required. An example of this direction is software like Linux Mint~\cite{linuxmint}, a Linux distribution that works ``Out of the Box'' but is still highly configurable, providing user-level control in an accessible way.

We focused on subsidiarity at the data collection stage: in the current technical landscape it is harder to control what a platform does with data once they have it, so collection is the first level of intervention. However, a more robust notion of subsidiarity would include the usage and retention stages of the data life cycle. User-level control of usage might look like Blockchain technologies that allow stakeholders to have finer-grained control over their data after it is collected~\cite{mann2021blockchain}. User-level control of retention would entail mechanisms that ensure data deletion~\cite{ng2019alexa}.

An emerging challenge to user-level control is the integration of LLM-based assistants in everyday applications, for example, the ``AI Overview'' in Google Search. There are pervasive public anxieties that LLMs are being trained on email and word processor data~\cite{westerholm2025, mccurdy2025}. While tech companies deny that they actually do this~\cite{westerholm2025, mccurdy2025}, these concerns highlight that more data is needed to train LLMs than is required for targeted advertising or general user experience improvements, thus exacerbating already-existing concerns with data collection. Although data collection for LLMs was an example mentioned by the interviewer in our thought exercises, our participants spoke more about targeted advertising. Yet, LLMs have relevance to many of the themes justifying user- or platform-level control; for example, platform-level control for improved user experience or user-level control for increased privacy. While our findings generally apply to the case of automatic data collection for LLM training, user-level control may be trickier to ensure in conversational interfaces: users may disclose more information to LLMs than on other platforms due to LLMs' conversational interface and imitation of human empathy~\cite{ali2025understanding}. While disclosing personal information to a conversational LLM technically falls under user-level control because the user is choosing to give up this data, is it really user-level control if the interface is subtly coercing them to do so?

\subsection{Trust in Communities Over Corporations}
Prior work found that willingness to share data can be dependent on who it is shared with~\cite{hudig2025intimate}. Building on this, our interviews revealed a greater trust in community organizations --- whether these are workplaces, schools, or special interest groups --- over ``big tech'' corporations. In many cases for our participants, the question was not \textit{if} they wanted their data controlled at a level higher than the user level, the question was \textit{who} is this higher level? When discussing community contexts, our participants highlighted that the community providing the software to them has a \textit{right} to control their data. However, they predominantly did not mention such a right when discussing the technology company that built the software, even though the software is, in a sense, being provided to them by the company. This underscores a difference in understanding of the rights of those \textit{developing} the technology and those \textit{administrating} the technology in terms of who has access to data. In addition to a different understanding of rights of access, there were also differing levels of trust in community organizations compared to tech companies. Our participants frequently mentioned a distrust of large tech companies, but also frequently mentioned they trust community-level institutions because of their different incentives and because they operate at a smaller scale. Additionally, a few participants mentioned a desire for something that doesn't currently exist: community collectives to help protect security or provide higher-level features typically provided by tech companies and powered by automatic data collection. Our participants want these higher-level features but are more comfortable sharing their data at the community level to receive them. We encourage this as a new direction.

\subsection{Limitations of Study and Future Work}
Data flow monitoring comes with a number of inherent limitations. Because the packets were encrypted, we were not able to read their contents. Thus, when we observed data transfers, we could not say for sure if the data was sensitive user data or not. We did our best to mitigate this by only considering larger packets, but the approach is not foolproof. Nevertheless, we hope that even if our data analysis approach is not absolutely precise, it still can provide a general sense of what is going on.

Another limitation lies in the software we studied, especially in certain categories. For example, in the chat application category we did not observe any applications that operated at a lower level of control, but our results may have been different if we included additional applications, such as Signal or Telegram. Additionally, although we only studied Windows and Linux because we could install these operating systems on laptops with the same hardware, Apple software products would provide an interesting comparison given their reputation of being more privacy-centric than other companies~\cite{leswing2021apple}. We recommend future work investigating more software platforms, especially chat applications and Apple products.

Finally, our interview sample consisted of graduate students and employees from our university. Thus, everyone in our participant group holds at least a bachelors degree, and many of our participants hold masters degrees or higher. While a higher level of education among our participants may have increased participants' familiarity with privacy concerns in technology and thus likely allowed them to give more informed responses, it also may have biased their responses. In particular, those with more privileged backgrounds do not suffer as many harms from data privacy violations as do members of marginalized groups~\cite{sannon2022privacy}. Thus, there is an important opportunity for future work exploring how people who are marginalized perceive what the ``necessary'' level of control is, as this will likely differ in comparison to the population we interviewed. Additionally, recruiting from our university meant that all interview participants currently reside in the Midwestern United States. While many of our participants were not originally from the Midwest, and some were not originally from the United States, there are still potential perspectives lost from this sample that would have surfaced in a more globally diverse interview pool. Given that perspectives of privacy can differ cross-culturally~\cite{trepte2017cross}, there is value in future work that investigates if cultural differences impact an understanding of which level of control is necessary in different technologies.

\section{Conclusion}
We proposed the principle of subsidiarity as a relevant critical lens through which to determine appropriate levels of control for software. We then applied the principle of subsidiarity to evaluate the tradeoffs between complete data privacy and automatic data collection in return for improved user experience. We did this by determining the level of control different everyday software platforms operate at through data flow monitoring, interviewing computer users to determine the level of control they find necessary for data on different types of software platforms, and comparing any mismatches in these findings. We found that our participants wanted user-level control for all software types, but were more willing to have platform-level control for browsers, search engines, and social media. In the case of group technology usage, our participants predominantly wanted control at the community level. We found the biggest violation of subsidiarity in chat applications: our participants wanted the greatest privacy in chat applications compared to the other software types, but none of the chat applications we studied offered a lower level of control. We encourage the development of more privacy-focused chat applications, and encourage the amount of privacy a platform provides to reflect the level of privacy users assume they will have based on its design. We additionally encourage greater control at the community level, which could help to bridge user desires for higher levels of control with distrust of large tech companies. Thus, the principle of subsidiarity is not just effective in considering how to bring greater autonomy back to the users, but also to creatively imagine new forms of governance that may better serve users in comparison to the current regime.


\bibliographystyle{ACM-Reference-Format}
\bibliography{sample-base}


\appendix
\section{Interview Guide}
The full interview protocol is as follows:

\textit{The interviewer first introduces the tradeoffs of giving control to higher or lower levels.}

Claimed Benefits:
Better understanding of user preferences to make a better product, better recommendation algorithms, allows for the product to be free, don’t have to make customization decisions (e.g. Apple’s design team may have better aesthetics than yours), able to share data with others for better cooperative work e.g. sharing calendar availability with coworkers, sharing location with friends and family.

Claimed Drawbacks:
Loss of control over one’s digital identity, ability for sensitive data to end up in the wrong hands, corporate or government surveillance including exploitation of user data for targeted advertising, behavioral modification, less ability to customize the experience to your liking (window managers, custom status bars, “ricing”) and pushing from the company to use their products (e.g. Microsoft strongly pushing people to use the Microsoft edge browser, Microsoft and Apple forcing people to use their AI tools in apps).

\textbf{Interviewer:} Rate on a scale of 1 to 5, 1 being not important at all that I have full control over my data, 5 being really important that I have full control over my data, how much do you want to control your data on different types of platforms?

\begin{itemize}
    \item Your computer’s operating system
    \item Your browser
    \item Search engines
    \item Email providers
    \item Social media platforms
    \item Chat platforms such as Slack or Discord
\end{itemize}

<Note: the data collected from these questions was redundant with data collected later in the interview and thus is not reported in the paper>

\textit{The interviewer then shares the findings from our data flow monitoring study with the participants.}

\textbf{Interviewer:} Is any of this surprising to you? Are you concerned about any of this?

\textit{The interview then proceeds to the thought exercises and identifications of necessary levels of control.}

\textbf{Interviewer:} Consider the \textbf{operating system of your laptop.}

Pretend you’re a software engineer at Microsoft working on the Windows OS team. Why might you think it’s necessary for Microsoft to control user data from the operating system?
\textit{Examples shared if the participant struggled:} crash reports/bug fixes, location data for weather services, personalized news offerings, collecting user data for product usage, collecting user data for AI training.

\textbf{Interviewer:} Pretend you are the organizer of a company, university, or special interest group that has provided laptops to all its members. Why might you think it’s necessary for you, the head of the organization, to control user data from the operating system?

\textit{Examples shared if the participant struggled:} software license compliance/automatic software updates, measuring how the laptops are benefitting the members or which software applications are being used the most in order to understand how the members of the organization are using the laptops and if they need extra training for certain types of software.

\textbf{Interviewer:} Why might it be necessary for all the data from your operating system to be completely controlled by you, the user?

\textit{Examples shared if the participant struggled:} privacy, especially because people likely keep a lot of sensitive information on their laptops.

\textbf{Interviewer:} Based on thinking through these different situations, what level do you think is simultaneously the lowest possible and highest level of control necessary for operating systems? Does this change in different contexts, e.g. personal or professional?

\textbf{Interviewer:} Consider a \textbf{web browser} like Google Chrome or Firefox.
Pretend that you’re a software engineer at Google, working on Google Chrome. Why might you think it would be necessary for Google to collect and control user data?
\textit{Examples shared if the participant struggled:} advertising to make the platform free, improved user experience.

\textbf{Interviewer:} Pretend that you are the head of a company, university or special interest group that has provided laptops to all its members. Why might it be necessary for you, the head of the organization, to control user data from the browser?

\textit{Examples shared if the participant struggled:} detecting phishing or guarding against unsafe downloads, compliance with group rules (e.g. no piracy or pornography), understanding how employees or members are using different web applications for targeted skill-building or digital literacy training

\textbf{Interviewer:} Why might it be necessary for a user to have full control over their browsing data?
\textit{Examples shared if the participant struggled:} data privacy – using the internet for personal reasons like looking up health information, advocating for oneself in a marginalized group, etc.

\textbf{Interviewer:} Based on thinking through these different situations, what level do you think is simultaneously the lowest possible and highest level of control necessary for web browsers? Does this change in different contexts, e.g. personal or professional?

\textbf{Interviewer:} Consider an \textbf{email platform} like Gmail.
Pretend that you’re a software engineer at Google on the Gmail team. Why might you think that it would be necessary for Google to collect and control user data and use it for purposes other than sending email?
\textit{Examples shared if the participant struggled:} spam filtering, taking flight data from email and turning it into google calendar events, targeted advertising to keep the platform free.

\textbf{Interviewer:} Pretend that you are the head of a company, university or special interest group that has provided laptops to all its members. Why might it be necessary for you, the head of the organization, to control user email data?

Examples shared if the participant struggled: phishing and scam detection, monitoring engagement with group communications (e.g. seeing if members open and read emails that are sent out), automatic meeting scheduling based on email contents.

\textbf{Interviewer:} Why might it be necessary that the individual user have full control over their email account?

\textit{Examples shared if the participant struggled:} data privacy, especially because emails can contain particularly sensitive data.

\textbf{Interviewer:} Based on thinking through these different situations, what level do you think is simultaneously the lowest possible and highest level of control necessary for email accounts? Does this change in different contexts, e.g. personal or professional?

\textbf{Interviewer:} Now consider a \textbf{search engine} like Google Search. Pretend that you’re a software engineer at Google. Why might you think that it would be necessary for Google to have control over user data?

\textit{Examples shared if the participant struggled: personalization for improved user experience, advertising that makes the platform free.}

\textbf{Interviewer:} Pretend that you are the head of a company, university or special interest group that has provided laptops to all its members. Why might it be necessary for you, the head of the organization, to control user data from a search engine?

\textit{Examples shared if the participant struggled}: blocking unsafe sites from search results, tailoring resources (if employees are searching the same problem there can be training sessions about it).

\textbf{Interviewer:} Why might it be necessary for the user to have full control over their search data?

\textit{Examples shared if the participant struggled:} data privacy --- people search for sensitive things.

\textbf{Interviewer:} Based on thinking through these different situations, what level do you think is simultaneously the lowest possible and highest level of control necessary for search engines? Does this change in different contexts, e.g. personal or professional?

\textbf{Interviewer:} Now consider a \textbf{social media platform} like Instagram, TikTok, YouTube, X, or Bluesky. Imagine you’re a software engineer for one of these platforms. Why might you think it is necessary to collect and control user data?

\textit{Examples shared if the participant struggled:} targeted advertising that keeps the platforms free, making recommendation algorithms more targeted to the user.

\textbf{Interviewer:} Currently, social media platforms like Instagram or X operate at the company level of control. However imagine a social platform for, say, a special interest group or hobby community, that is controlled entirely by the leader of the group or another community admin. If you are the admin for this special interest group's platform, why might it be necessary for you to control the group’s data?

\textit{Examples shared if the participant struggled:} community moderation and safety, program engagement (see which hash tags, topics, or accounts users interact with the most to better tailor community discussions).

\textbf{Interviewer:} Why might it be necessary for a user to have full control over their social media data?

\textit{Examples shared if the participant struggled:} data privacy.

\textbf{Interviewer:} Based on thinking through these different situations, what level do you think is simultaneously the lowest possible and highest level of control necessary for social media platforms? Does this change in different contexts, e.g. personal or professional?

\textbf{Interviewer:} Now consider a \textbf{chat platform} like Slack, Discord, Whatsapp, or iMessage. Imagine you’re a software engineer for one of these platforms. Why might you think it’s necessary to collect and control user data?

\textit{Examples shared if the participant struggled:} selling data to advertisers to keep the platform free, crash reports.

\textbf{Interviewer:} Imagine you are the head of a company, university, or special interest group that uses a chat platform like Slack or Discord for its internal communications. Why might it be necessary for you to control the group’s data?

\textit{Examples shared if the participant struggled:} rule enforcement (e.g. anti harassment) and conflict resolution.

\textbf{Interviewer:} Why might it be necessary for a user to have full control over their chat platform data?

\textit{Examples shared if the participant struggled:} data privacy, especially of sensitive information or personal conversations.

\textbf{Interviewer:} Based on thinking through these different situations, what level do you think is simultaneously the lowest possible and highest level of control necessary for chat platforms? Does this change in different contexts, e.g. personal or professional?

\end{document}